%% file: main.tex
\documentclass{aa}
\usepackage{graphicx}
\usepackage[utf8]{inputenc}
\usepackage{textcomp}
\usepackage{gensymb}
\usepackage{xcolor}
\usepackage{txfonts}
\usepackage{tabularx}
\usepackage{amsmath}
\usepackage{amsfonts}
\usepackage{amssymb}
\usepackage{mathrsfs}
\usepackage{orcidlink}
\usepackage{hyperref}
\usepackage{breakcites}
\usepackage{physics}
\usepackage{booktabs}
\usepackage{multirow}
\usepackage{soul}
\newcommand*{\Lw}{\textit{Lightweaver}}
\newcommand*{\MpiAmr}{\texttt{MPI-AMRVAC}}
\newcommand*{\Caii}[1]{Ca~\textsc{ii}#1}
\newcommand*{\Mgii}[1]{Mg~\textsc{ii}#1}
\newcommand*{\halpha}{H$\alpha$}
\defcitealias{Heinzel:2015}{Hea15}

\input{acronyms}

\begin{document}
	
\title{1.5D NLTE spectral synthesis of a 3D filament/prominence simulation}

\author{J. M. Jenkins
      \inst{\ref{KUL}}\orcidlink{0000-0002-8975-812X}
      \and
      C.\,M.\,J. Osborne
      \inst{\ref{Glasgow}}\orcidlink{0000-0002-2299-2800}
      \and
      R. Keppens
      \inst{\ref{KUL}}\orcidlink{0000-0003-3544-2733}
      }

\institute{Centre for mathematical Plasma-Astrophysics, Celestijnenlaan 200B, 3001 Leuven, KU Leuven, Belgium \label{KUL}\\
    \email{jack.jenkins@kuleuven.be}
\and
    SUPA School of Physics and Astronomy, University of Glasgow, Glasgow G12 8QQ, UK \label{Glasgow}
    }

\date{Received September 2022; accepted November 2022}

 
\abstract
{Overly idealised representations of solar filaments/prominences in numerical simulations long limited their morphological comparison against observations. Moreover, it is intrinsically difficult to convert simulation quantities into emergent intensity of characteristic, optically-thick line cores and/or spectra that are commonly selected for observational study.}
%
{We here demonstrate how the recently developed \Lw{} framework makes non-\acs{LTE} (\acs{NLTE}) spectral synthesis feasible on a new 3D \textit{ab-initio} magnetohydrodynamic (MHD) filament/prominence simulation, in a post-processing step.}
{We clarify the need to introduce filament/prominent-specific \Lw{} boundary conditions that accurately model incident chromospheric radiation, and include a self-consistent and smoothly varying limb darkening function.}
{Progressing from isothermal/isobaric models to the self-consistently generated stratifications within a fully 3D MHD filament/prominence simulation, we find excellent agreement between our 1.5D \ac{NLTE} \Lw{} synthesis and a popular Hydrogen~H$\alpha$ proxy. We compute additional lines including \Caii{~8542} alongside the more optically-thick \Caii{~H\&K}~\&~\Mgii{~h\&k} lines, for which no comparable proxy exists, and explore their formation properties within filament/prominence atmospheres.}
{The versatility of the \Lw{} framework is demonstrated with this extension to 1.5D filament/prominence models, where each vertical column of the instantaneous 3D MHD state is spectrally analysed separately, without accounting for (important) multi-dimensional radiative effects. The general agreement found in the line core contrast of both observations and the \Lw{}-synthesised simulation further validates the current generation of solar filaments/prominences models constructed numerically with \MpiAmr{}.}



\maketitle

\section{Introduction}
Solar prominences and filaments are clouds of kK plasma, oftentimes referred to as `chromospheric', suspended within and thermally isolated from the ambient MK solar corona. 
Prominences appear in observations projected above the solar limb whereas filaments appear projected against the solar disk. 
As identical phenomena, despite the two names, their difference results solely from their projection as the solar surface rotates from the perspective of the Earth.
These structures are observed to form, evolve, and dissipate over timescales ranging from days to months, all the while displaying a wide range of internal dynamics with lifetimes on the order of minutes to hours \citep[][]{Labrosse:2010, Mackay:2010}.
Should their bounding magnetic topology lose equilibrium, a global eruption can lead to their embedding within coronal mass ejections with implications on the near-Earth environment \citep[][]{Vial:2015}.

Despite routine observations over many decades, the diagnosing of plasma conditions within solar prominences/filaments continues to suffer from observational restrictions associated with spatial, spectral, and temporal resolution; the common approach being to maximise two at a cost for the third \citep[even for our most state-of-the-art models \textit{e.g.},][]{Levens:2016a, Levens:2016b, Peat:2021}. 
On the other hand, numerical models of solar prominences and filaments have advanced significantly within the last decade \citep[\textit{e.g.},][]{Hillier:2011, Hillier:2013, Khomenko:2014, Xia:2014, Terradas:2015a, Terradas:2015b, Xia:2016b, Kaneko:2018, PopescuBraileanu:2021a, PopescuBraileanu:2021b}.

As recently demonstrated by \citet{Jenkins:2022}, the increasing complexity of solar prominence/filament models is rapidly closing the resolution gap between numerical simulations and equivalent observations. 
These authors and numerous others validated their simulations against observations by converting the primitive variables of their numerical model to integrated intensity quantities that mimic the optically-thin coronal \ac{EUV} observations of the \ac{AIA} on board the \ac{SDO}. 
However, filaments and prominences have non-negligible optical thicknesses, appearing in absorption in \ac{AIA} observations due to scattering photoionisation by the Hydrogen Lyman continuum \citep[][]{Kucera:1998,Williams:2013}. 
The cooler, optical lines formed within solar prominences/filaments then have much larger optical thicknesses than for the \ac{EUV} case \citep[][]{Anzer:2005}. 
As optical thickness increases, the encoding of information within the emergent intensity loses the simple assumption of a 1-1 translation from the local properties of primitive (pressure, density, temperature, etc.) variables, depending instead on the nonlocal and noninstantaneous state of the atmosphere \citep[][]{Rutten:2019}. 
For the synthesis of the optical Hydrogen H$\alpha$ line, \citet{Jenkins:2022} employed the approximate method presented by \citet{Heinzel:2015} (hereafter \citetalias{Heinzel:2015}). 
Even for such a line core that tends to straddle the divide between optically-thin and optically-thick behaviour, the tables of \citetalias{Heinzel:2015} facilitate the conversion of the aforementioned simulation quantities using a series of approximately linear relationships. 
The subsequent matching of features between their simulation and observations in \citet{Jenkins:2022} suggests the applicability of such an approximate synthesis method \citep[see also][]{Gunar:2016b, Gunar:2018, Jenkins:2021}. 
For very optically-thick lines, however, such a simple mapping is not possible and instead models that consider the departure from \ac{LTE} i.e., non-\ac{LTE} (hereafter \ac{NLTE}), are required to more-accurately represent the multi-dimensional, nonlocal, and perhaps temporally-dependent matter-light interaction \citep[][]{Labrosse:2016}.

Efforts to model the emergent spectra of moderately optically-thick lines sourced within prominence and filaments began with somewhat idealised isothermal/isobaric slab/thread models \citep[][]{Gouttebroze:1993}, eventually progressing to include the sharp temperature transition of the \ac{PCTR} necessary to accurately synthesise the very optically thick lines of Hydrogen and Magnesium~{\sc ii} \citep[][]{Heinzel:2001, Heinzel:2014}. Until recently, these models have focused on the intricate dependence of emergent intensity on a range of stratified 1.5\,--\,2.5D atmospheres \citep[\textit{e.g.},][]{Heinzel:2014, Labrosse:2016, Levens:2019}. Even for such idealised stratified atmospheres, there is degeneracy when inverting from only the shape of the associated spectra, and this hampers the use of more advanced models \citep[current efforts to minimise this are expanding to include t-distributed stochastic neighbouring, \citet{Meetu:2021}, and principle component analysis,][]{Dineva:2020}. \citet{Gunar:2015} took a more consistent approach by constructing model threads under the \ac{MHS} assumption according to a magnetic arcade topology derived from \ac{NLFFF} extrapolations \citep[][]{Gunar:2016a, Gunar:2016b, Gunar:2018,Gunar:2019}. Nevertheless, these authors then also applied the approximate radiative transfer modelling approach of \citetalias{Heinzel:2015}. 

Running parallel to this, multiple authors have constructed numerical models of the solar chromosphere and progressed successfully from 1.5D \ac{NLTE} statistical equilibrium and radiative transfer calculations to a full 3D synthesis \citep[commonly referred to as \ac{RMHD} simulations][wherein tabulated radiative losses are coupled to the base \ac{MHD} state]{Carlsson:1986, Carlsson:1997, Leenaarts:2007, Leenaarts:2012a, Bjorgen:2018}. Such a dedicated effort to model a 1.5 and 3D chromosphere led to the early development of models that yielded accurate spectra, in addition to 2D maps that mimic narrowband imagery of modern telescopes, most recently with \citet{Bjorgen:2019}. Solar filaments and prominences, on the other hand, have historically suffered from a lack of self-consistent, dynamic models to which one may apply the equivalent synthesis and associated analysis (\citet{Heinzel:2006}, with the closest being the aforementioned \ac{MHS} case of \citealt{Gunar:2015}). With the recent development of a suitable, state-of-the-art model according to \citet{Jenkins:2022}, we thus present the first steps towards applying similar modelling, as that of the chromosphere, to solar filaments/prominences using the new \Lw{} framework.

In Section~\ref{s:methods} we outline the \Lw{} framework, and detail the addition of new boundary conditions simultaneously suitable for both prominence and filament atmospheres. In Section~\ref{s:Results} we present the syntheses resulting from applying \Lw{} to a 3D, nonadiabatic \ac{MHD} simulation of a prominence/filament. We compare, contrast, and summarise the relevant diagnostics and associated limitations in Section~\ref{s:Discussion}, before closing with a summary of the anticipated next steps in Section~\ref{s:summary_and_conclusions}.



\section{Methods} \label{s:methods}
In \citet{Jenkins:2021,Jenkins:2022}, we used the \texttt{MPI-AMRVAC 2.0} toolkit \citep[][]{Xia:2018,Keppens:2021} to construct realistic 2.5~\&~3~dimensional representations of solar prominences and filaments \citep[see also,][]{Kaneko:2018}. 
This was corroborated through a direct comparison between simulations and observations. Specifically, the application of a combination of \ac{EUV} and Hydrogen-H$\alpha$ proxies were shown to yield imagery that resembled the appearance of prominences and filaments within the actual solar atmosphere. 
Unfortunately, there exists only a limited number of these proxies available through which we can compare simulations against observations; prominence and filament studies are particularly limited as a consequence of there being only a few lines within which they are visible. 
Furthermore, these representations are built on a number of assumptions that neglect a significant amount of information.
The perhaps most crucial of which being the influence of instantaneous dynamics \textit{i.e.}, flows within the plasma. 

\subsection{The \Lw{} Framework} \label{ss:lightweaver}
The recently developed \Lw{} framework \citep[][]{Osborne:2021} is used to solve the radiative transfer equation and statistical equilibrium equations for a given stratification of atmospheric parameters.
\Lw{} determines the non \ac{LTE} populations of the species in the plasma by iteratively computing the associated radiation field (using the cubic Bézier short characteristic formal solver of \citet{DelaCruzRodriguez2013}) and then updating the atomic level populations taking into account the updated radiative and collisional rates (using the fully preconditioned \ac{MALI} method \citep{Rybicki1992, Uitenbroek2001}).
The alternating iteration of these two steps continues until the maximum relative change of the atomic level populations falls below an arbitrary threshold indicating convergence (in our case the typical $10^{-3}$), and the maximum relative change in angle-averaged intensity at each frequency and location in the model falls below $3\times10^{-3}$.
The use of the fully preconditioned MALI technique ensures that photoionisation interactions between the \ac{NLTE} species are considered, for instance both the hydrogen Lyman continuum and lines can have a significant effect on the Ca\,\textsc{ii} level populations and line shapes \citep[e.g.][]{Ishizawa1971, Gouttebroze:2002}. Dynamic electron/ionisation equilibrium (charge conservation) is not considered here, fixing $n_\mathrm{e}$ instead according to the tables of \citetalias{Heinzel:2015} and the initial atmospheric stratification.

Some spectral lines, especially strong resonance lines that form in regions where radiative effects dominate over collisional effects are affected by \ac{PRD}, where the absorption and emission frequency of a photon is correlated across the spectral line.
Normally, the assumption of \ac{CRD} is made, whereby the line emission profile is the same as the absorption profile.
For the simulations presented here, we consider the hydrogen, calcium, and magnesium populations outside of \ac{LTE}, as such the Ly~$\alpha$\&$\beta$, Ca~\textsc{ii}~H\&K, and Mg~\textsc{ii}~h\&k lines are treated with \ac{PRD} \citep[][]{Paletou:1993}.
To compute the line emission ratio, \Lw{} adopts the iterative method described in \citet{Uitenbroek2001}, and updates this after every population update.
The anisotropy of the angle-averaged radiation field due to plasma flows are accounted for using the hybrid method of \citet{Leenaarts:2012b}, which interpolates the angle-averaged radiation field to the plasma rest frame to compute the line emission ratio. For all models shown in this paper, the \texttt{v0.8} release of \Lw{} was used \citep{LwV08}\footnote{\url{https://doi.org/10.5281/zenodo.6598463}}.



\subsubsection{The 1.5D geometry approximation, implementation, and associated limitations} \label{sss:1.5D_geom}

\begin{figure}
    \centering
    \resizebox{\hsize}{!}{\includegraphics[clip=,trim=280 0 70 0]{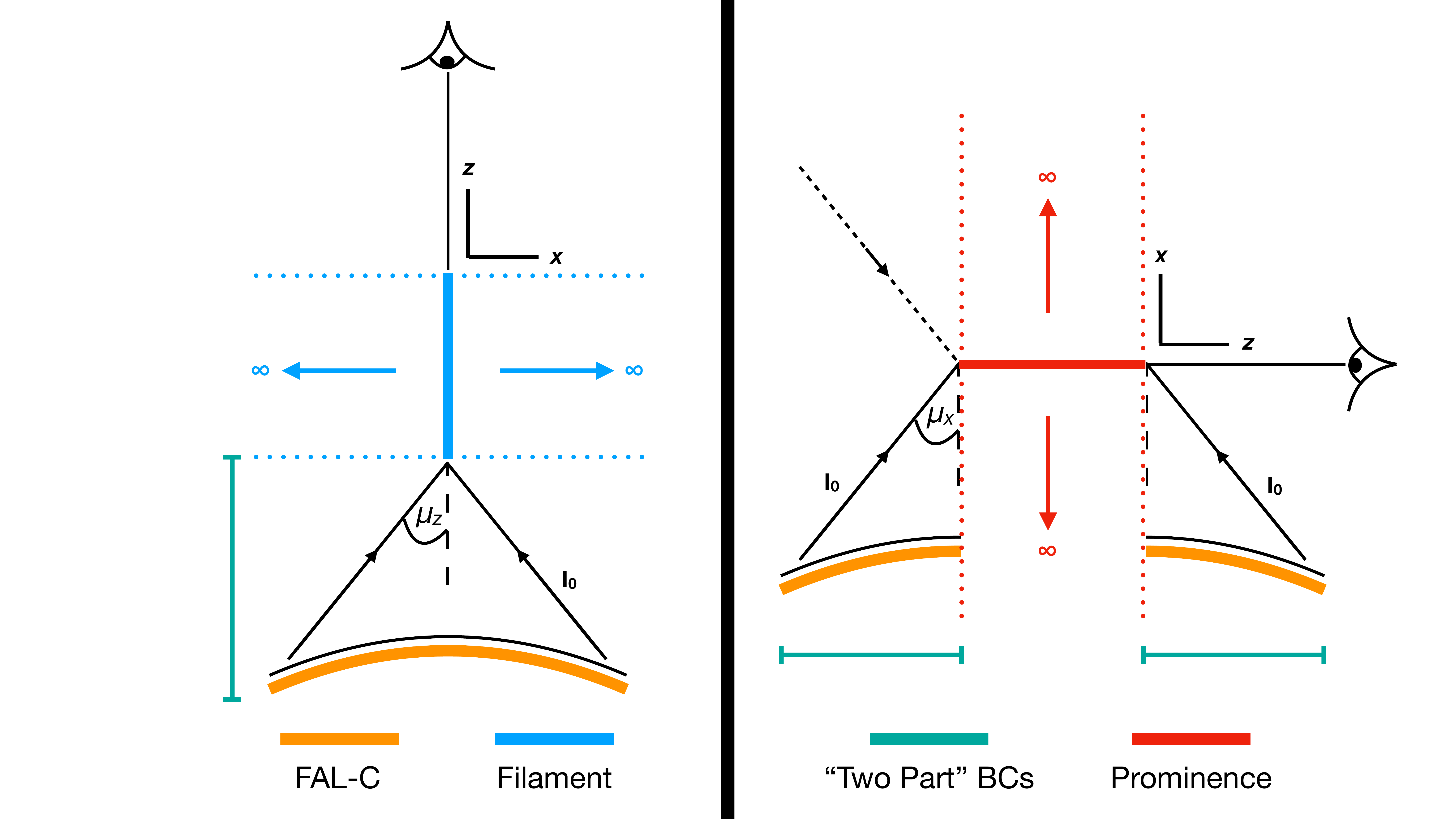}}
    \caption{The 1.5D geometry assumed for the filament and prominence models. An angle-dependent FAL-C semi-empirical chromospheric model plus diffused Planck function is used to construct the wavelength and $\mu$-dependent radiation field input into the filament/prominence stratifications. The prominence case adopts a rotated reference frame to account for the differing projection. Regions comprising the `two part' boundaries are indicated. Coronal illumination is ignored in all cases.}
    \label{fig:problem_geometry}
\end{figure}

\begin{figure*}
    \centering
    \resizebox{0.9\hsize}{!}{\includegraphics{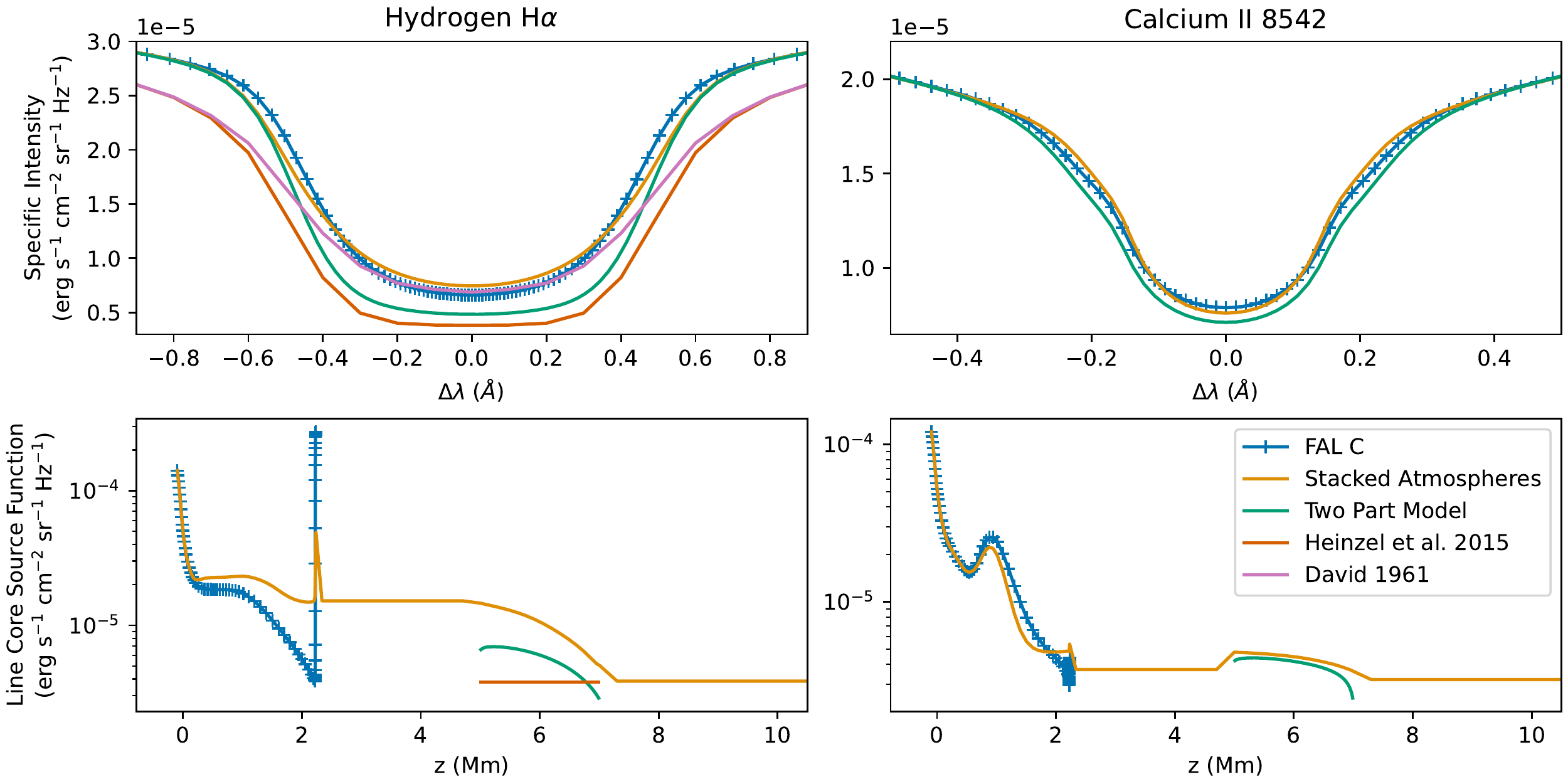}}
    \caption{A comparison between the \Lw{} synthesis of the FAL-C, stacked, and two-part models. \textit{Top row}; The synthetic spectra for Hydrogen H$\alpha$ and the Calcium II 8542~\AA\ of the IR triplet for each of the models. For reference, the two-part model was also synthesised using the method of \citetalias{Heinzel:2015}, along with the disk centre spectrum of \citet{David:1961} which is used as its associated background intensity. \textit{Bottom row}; The stratified source functions corresponding to the spectra synthesised in the top row.}\label{fig:15d_vs_2part}
\end{figure*}

\Lw{} currently supports plane-parallel and two-dimensional Cartesian descriptions of the atmospheric parameters.
For this work, we will restrict ourselves to the 1.5D geometry assumption when considering our atmospheres. Comparisons between this baseline and an extension to higher dimensionality are reserved for a subsequent study. Hence, we will treat every vertical column of our 3D simulated stratification as being a local 1D stratification that is geometrically invariant in the other two dimensions, the so-called `plane-parallel' approximation. 
This means that each column is considered completely independent of its surroundings \citep[cf.][]{Leenaarts:2012a}. 
In this work, we will focus solely on the vertical and horizontal axis-aligned projections for the filament and prominence cases, respectively, the geometry for which is shown graphically in Figure~\ref{fig:problem_geometry}. 
Our initial tests indicated such a 1.5D approximation to have important consequences on the shape of the spectra, in particular for those synthesised for the filament projection. 
As any given stratification within the \ac{MHD} model contains multiple strong gradients, in particular for the \ac{LOS} velocity that is ordinarily omitted, we will demonstrate the limitations, as briefly acknowledged by \citet{Paletou:1993}, with the use of a far simpler isothermal, isobaric model (with fixed ionisation degree). 
In this isothermal-isobaric model the filament is simply a static, extended region with plasma parameters taken from the FAL-C semi-empirical atmosphere at a temperature of 8635~K. 
We will then go on to detail how we overcame these issues.

A filament projection is by definition observed against the bright background of the solar disk; spectral observations of filaments describe an absorption signature imposed on a background containing a `continuum' and the `average chromospheric profile'. 
The continuum component is typically sourced from photospheric heights where we may assume \ac{LTE} and thus the wavelength-dependent blackbody spectrum \textit{i.e.}, the Planck function. 
For the chromospheric component of the 1.5D filament stratification, we choose the FAL-C model of \citet{Fontenla:1993} that spans from 100~km below the base of the photosphere up to a transition-region height of $\approx$~2.2~Mm, encompassing the chromosphere in between. 
It is on top of this `base' that we then stack the isothermal, isobaric filament atmosphere. 
The upper boundary is assumed to be \textit{open}, meaning we neglect all (EUV) radiation incident from the corona \citep[as would be required to consider the He~{\sc i}~10830 triplet state also commonly used to study filaments \textit{e.g.},][]{Labrosse:2016}.

\begin{figure*}
    \centering
    \resizebox{0.9\hsize}{!}{\includegraphics{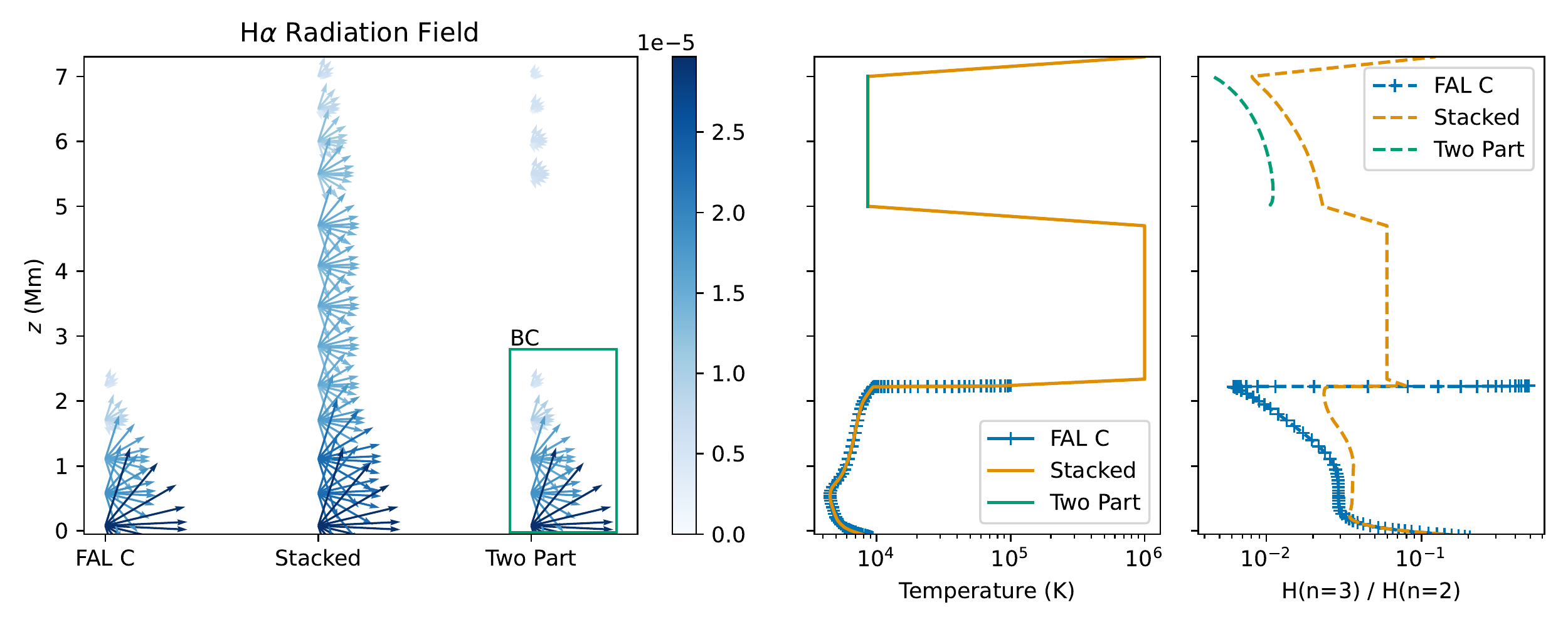}}
    \caption{The influence of the stratification structure on the internal radiation field. The anisotropy of the radiation field in the line core ($\Delta \lambda = 0$) of H$\alpha$ is shown in the left-hand panel. Here, the arrows represent the $\mu$ angles considered for the internal energy balance, and their length/color their corresponding amplitude. The green box around the lower region of the Two Part model indicates that this contribution is contained within the boundary condition. On the right the associated temperature stratification (solid lines) of the atmospheres and ratio of populations responsible for forming H$\alpha$ (dashed lines) are shown.}\label{fig:15d_aniso_temperature}
\end{figure*}

As we show in Figure~\ref{fig:15d_vs_2part}, for the converged (cf. Section~\ref{ss:lightweaver}) stacked-atmosphere (orange) temperature profile (shown in the right panel of Figure~\ref{fig:15d_aniso_temperature}), the resulting spectra characterises the filament line core signature of Hydrogen~H$\alpha$ as having a positive contrast compared to an isolated FAL-C (blue crosses) atmosphere. 
This is in complete contradiction to observational conclusions, and we find there to be two primary (coupled) reasons for this. 
First, the incident radiation entering the lower boundary of the filament is diluted but not limb darkened; the infinitely-wide plane-parallel 1.5D approximation geometrically prohibits such a consideration.
Second, and more crucially, radiation released from lower altitudes can be \textit{trapped} in the region between the chromosphere and the filament. When the filament is modelled in this 1.5D way, it has infinite horizontal extent, and all energy leaving the chromospheric model has to pass through it. 
This radiation is then absorbed by the filament, after which the excited populations will spontaneously decay -- by definition releasing radiation isotropically -- and hence direct a significant portion of this energy back towards the chromosphere. 
The iterative solution will then effectively `pump' the region between the chromosphere and filament with this additional down-going radiation.


The left panel of Figure~\ref{fig:15d_aniso_temperature} provides a graphical representation of this `radiation trapping'. The arrows represent the radiation along $\mu$ angles considered for the energy transport throughout the stratifications. 
In comparison with the FAL-C atmosphere, the H$\alpha$ line core ($\Delta \lambda = 0$) radiation field in the stacked model does not fall off at the top of the chromosphere ($\approx$~2.2~Mm), as it should, when the two atmospheres are combined in this way. 
Furthermore, the ratio of Hydrogen $n=3$ to $n=2$ population levels shown in the right panel of the same figure contains a clear enhancement throughout the atmosphere, in particular for the chromosphere which \textit{should} ordinarily be defined by the FAL-C properties alone with minimal influence anticipated from the filament. 

To appreciate the influence of this enhancement on the emergent intensity, we introduce the frequency $\nu$ and angle $\mu$ dependent \textit{formal solution} to the \ac{RTE},
\begin{equation}
    I_{\nu,\mu}(\tau_{\nu,\mu}) = I_{\nu,\mu}(0)\,e^{-\tau_{\nu,\mu}} + \int_0^{\tau_{\nu,\mu}} S_{\nu,\mu}(\tau'_{\nu,\mu})\,e^{-(\tau_{\nu,\mu}-\tau'_{\nu,\mu})}\,
    d\tau'_{\nu,\mu}, \label{eq:formal_RTE}
\end{equation}
also commonly referred to as the transport equation in integral form. Here, $\tau_{\nu,\mu}$ is the optical depth (thickness) given by,
\begin{equation}
    \tau_{\nu,\mu} = \int^{s_2}_{s_1} \alpha_{\nu} \frac{ds}{\mu}, \label{eq:optical_depth}
\end{equation}
for which $\alpha_\nu$ is the absorption coefficient and hence Eq.~\ref{eq:optical_depth} describes the total absorption encountered by a ray passing through some material of length $ds$. 
With this, we see that Eq.~\ref{eq:formal_RTE} considers some initial incident radiation (from the solar surface, for example) $I_{\nu,\mu}(0)$ attenuated by the total absorption of the atmosphere under consideration, and the integrated contribution of the continuous local emission or absorption properties $S_\nu$ of the atmosphere that are at each \textit{point} further attenuated by the absorption of the remaining atmosphere.


The source function $S_\nu$ is approximately proportional to the ratio of the upper $n_u$ to lower $n_l$ population levels $n_u/n_l$ for a given transition.
From the right panel of Figure~\ref{fig:15d_aniso_temperature} we already found this ratio to be enhanced throughout the atmosphere, including the chromosphere, as a consequence of the aforementioned pumping. 
This trapping is therefore responsible for the enhancement in the H$\alpha$ source function within the chromospheric component of the model (lower-left panel of Figure~\ref{fig:15d_vs_2part}), and for in turn driving the line into emission relative to the reference FAL-C chromosphere.
The filament itself is not solely responsible for the enhanced profiles, as it is the \textit{combination} of the chromosphere with enhanced source function and the response of the atomic populations within the filament that are responsible \cite[see the reference to unpublished computations in][]{Paletou:1993}.
In observations, typical filaments are characterised as having relatively thin yet extended aspects, and so the invariance assumption is approximately valid for one of the dimensions, and so much of the $\mu$ space should \textit{not} encounter this reflective radiation trapping property that we find here \citep{Labrosse:2010, Mackay:2010, Parenti:2014, Vial:2015}. 
Much of the chromospheric radiation should instead `free-stream' out of the local volume.

For both of these reasons, we adopt the two-part model that will be described in Section~\ref{sss:two-part_geom}, and treat the chromosphere as a radiative boundary condition to the filament. 
This also enables prominence synthesis, as 1D plane-parallel models that need to be stacked on a chromosphere cannot be used for prominence modelling \citep[\textit{e.g.},][as the models are infinite along such a \ac{LOS}]{Paletou:1993}. 
A comparison of these two methods will then be presented in Section \ref{sss:15d_two-part_comparison}.

\subsubsection{Two-part slab model} \label{sss:two-part_geom}

The \textit{default} boundary conditions for the MALI approach implemented within \Lw{} assume a `ThermalisedRadiation' (diffused Planck function) at the lower boundary, and a `ZeroRadiation' (open) case for the top of the plane-parallel atmosphere. 
Considering we wish to preserve the average profile of the FAL-C model, we can move this portion of the stratification into the boundary condition in combination with the `ThermalisedRadiation' condition. Furthermore, it was noted how the limb darkening effects on the radiation incident on the underside of the filament atmosphere cannot be directly considered for the stacked plane-parallel case.
Hence, we can instead compute the emergent specific intensity from the combined `ThermalisedRadiation' + FAL-C atmosphere for a range of $\mu$ \textit{a-priori}, adopt this as the lower boundary condition, and feed this directly into a purely corona+filament stratification.
In this way, those incident rays that approach the solar limb are self-consistently darkened, infinitely so if they do not encounter the solar limb at all.
The approach of modelling the filament as an isolated structure with a boundary condition that describes the incident radiation is already the standard approach in both plane-parallel isothermal/isobaric and \ac{PCTR} filament modelling \citep[e.g.][]{Gouttebroze:1993, Heinzel1995, Paletou1995, Heinzel:2014}.
Our method then represents an important addition to this existing state-of-the-art by considering self-consistent atmospheric stratifications that include detailed velocity profiles, synthesised with an additional detailed angular variation in the radiation incident on the bottom of the filament (geometric limb darkening).

By default, the boundary conditions in \Lw{} are both fully angle- and wavelength-dependent so as to enable an accurate treatment in those situations where the incident radiation may be anisotropic (\textit{e.g.}, in the presence of strong flows or when modelling an eruptive process).
We approach this in much the same way as \cite{Gouttebroze:2005} and their follow-on studies by considering the incident radiation for each discrete ray as an angular average through the opening angle of a series of nested cones, the explicit geometry considerations for which can be found in Appendix~\ref{a:boundaries}.
The difference herein being the specific intensity incident on the bottom of our filament atmospheres is instead self-consistently computed using a unified technique across the entire spectral range under consideration. That is to say, it is not dependent on an ad-hoc treatment of any observational spectra, nor any assumed or fitted limb darkening functions, be them fixed to set wavelengths/line cores or across limited wavelength ranges \citep[as is the standard approach cf.][and numerous others]{Gouttebroze:1993,Paletou:1996,Gouttebroze:2002,Gouttebroze:2004,Gouttebroze:2005,Gouttebroze:2006,Gouttebroze:2007,Leger:2007,Gouttebroze:2008, Leger:2009}. The result is a fully consistent model across all angles, wavelengths, and transitions based on the underlying plane-parallel FAL-C model.

This approach then has the additional operational advantage that the equilibrium within the lower FAL-C portion of the atmosphere need not be dynamically considered for each column, instead computed once and stored meaning numerical convergence times are significantly reduced.

\subsubsection{Comparison of the treatments}\label{sss:15d_two-part_comparison}

\begin{figure*}
    \centering
    \resizebox{\hsize}{!}{\includegraphics{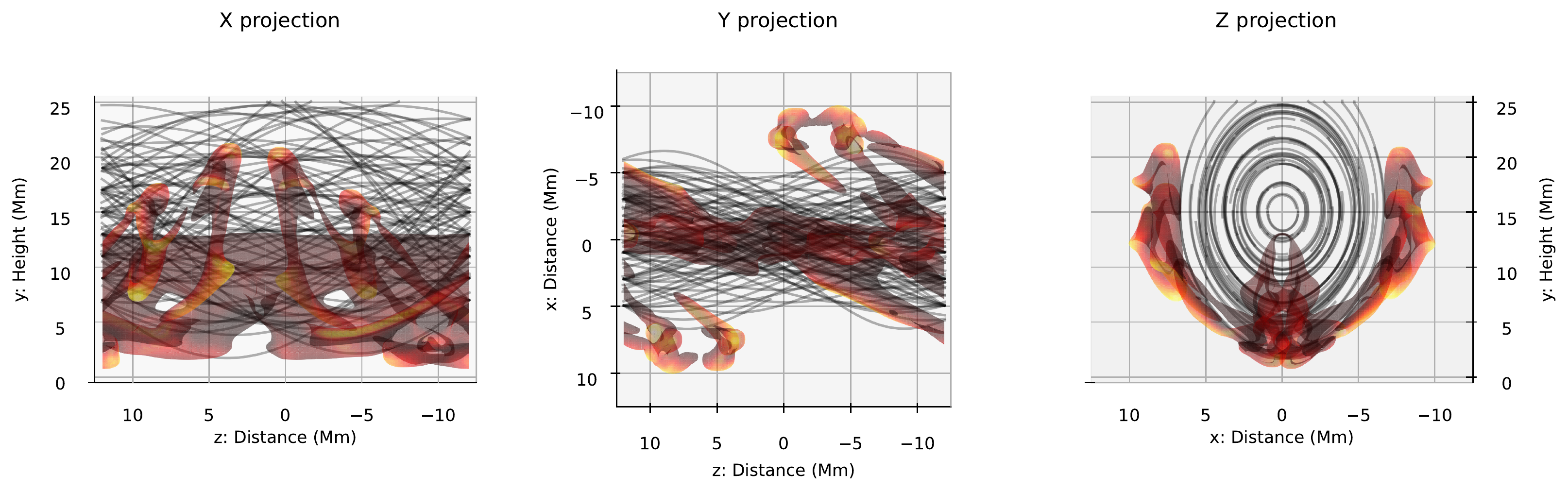}}
    \resizebox{1.0\hsize}{!}{\includegraphics{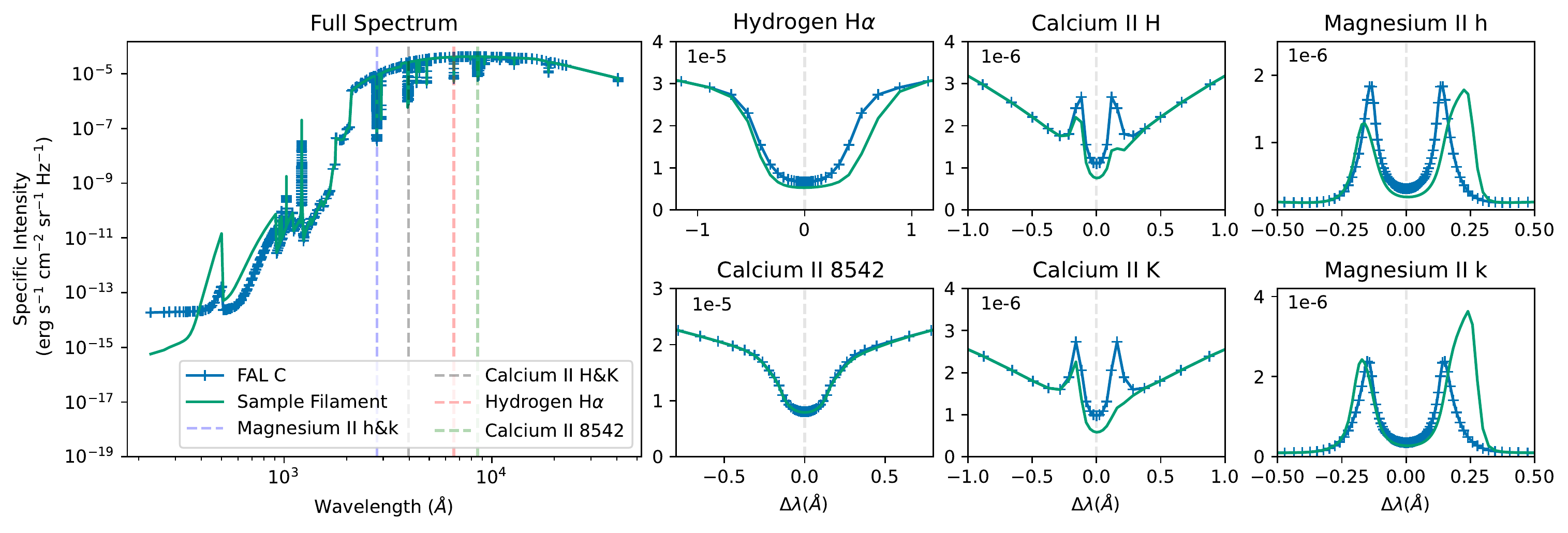}}
    \caption{\textit{Top row}; Axis-aligned representations of the solar filament/prominence simulation completed using \MpiAmr{}. The cool plasma is bound by a semi-transparent density isocontour of value 1~$\times$~10$^{-14}$~g~cm$^{-3}$ and coloured black-red-yellow according to the local temperature, cold to hot respectively. Magnetic field lines that reveal the bounding flux rope are traced and overlaid using black, semi-transparent lines. \textit{Bottom row}; Comparison between the synthesis of a coronal, FAL-C model and that of a central position (0,0)~Mm within the `Y projection'. On the right we show a sample of six lines typically used to observe filaments: \halpha{}, \Caii{~8542}, \Caii{~H\&K}, and \Mgii{~h\&k}.}
    \label{fig:topology_example_filament_spectra}
\end{figure*}

The spectra synthesised from the example isothermal and isobaric filament atmosphere using the modified, two-part model (green) is also shown in Figure~\ref{fig:15d_vs_2part} for which a clear negative contrast is now present at $\Delta \lambda = 0$.
Comparing the line core source functions, we now see that not only was the chromosphere heavily influenced by the stacked atmosphere construction, but also the filament itself.
The two-part model construction then appears to overcome this problem, in particular for the case of the Hydrogen H$\alpha$ line. 
Such a specific shape for the variation of $S_\nu$ with height now qualitatively traces that previously calculated for the \textit{bright rim} prominence phenomenon \citep[cf. Figure~5 of][]{Heinzel:1995}.

Figure~\ref{fig:15d_vs_2part} also shows H$\alpha$ synthesis using the approximate method of \citetalias{Heinzel:2015}, along with the associated disk-centre reference spectrum of \citet{David:1961}. 
We note that this reference spectrum agrees well with the FAL-C synthesis in the line-core, but does not rise up to the continuum level as quickly.
Nevertheless, for the far wings of H$\alpha$ ($\Delta\lambda \ge10\AA$), the FAL-C and reference spectrum agree very well once again.
This could likely be improved with a more accurate treatment of resonance broadening (Heinzel, priv. comm).
The approximate method of \citetalias{Heinzel:2015} synthesises the line-core intensity of H$\alpha$ assuming a \ac{CRD} formalism (whereby the source function is constant across the line), and a Voigt absorption profile with the same damping terms as used for the two-part model.
This H$\alpha$ proxy has a very similar shape to the two-part \Lw{} model, albeit with a deeper line-core, and deeper wings inherited from the reference incident spectrum \citep[][]{David:1961}.
Thus, for the simple isothermal and isobaric model, this proxy and the two-part model both produce spectra with the expected shape and very comparable forms.

\subsection{Application to a \ac{MHD} model of a solar filament/prominence} \label{ss:mhd_filprom}

In the recent study of \citet{Jenkins:2022} we presented a fully 3D filament/prominence model constructed \textit{ab-initio} following the `levitation-condensation' formation mechanism. 
For this study, we have reduced the resolution of the simulation domain in comparison to the one that was presented in that study. 
Hence, the simulation domain spans $-12 < x < 12$, $-12 < z < 12$~Mm in the horizontal, and $0 < y < 25$~Mm in the vertical with a uniform base resolution of 144$^3$ grid cells, each of physical $x,y,z$ dimensions $167~\times~173~\times~167$~km. 
All other settings are identical to the simulation described in \citet{Jenkins:2022}, wherein the associated numerical algorithmic details can also be found.

The levitation-condensation process within this \ac{MHD} simulation follows the identical evolution as in \citet{Jenkins:2021, Jenkins:2022}, that is, an initial linear force-free magnetic field configuration is deformed following driving motions imposed within the bottom boundary conditions. 
This drives the footpoints of the magnetic field towards $x=0$ whereby reconnection initiates and drives the construction of the coronal flux rope. 
The material within this flux rope is then isolated from the heat flux supplied from the bottom of the simulation domain by field-aligned thermal conduction, and is free to cool. 
Upon the triggering of the thermal instability, discrete condensations begin to form and slide down magnetic field lines due to gravity \textit{i.e.}, towards lower heights. 
At $t\approx$~6400~s, the flux rope and filamentary condensations have formed, of which some have settled in topological magnetic dips whilst others are still falling. 
The top row of Figure~\ref{fig:topology_example_filament_spectra} presents an isocontour and fieldline representation of the simulation domain at this time, as viewed projected along the three coordinate axes.

\begin{figure*}
    \centering
    \resizebox{1.\hsize}{!}{\includegraphics{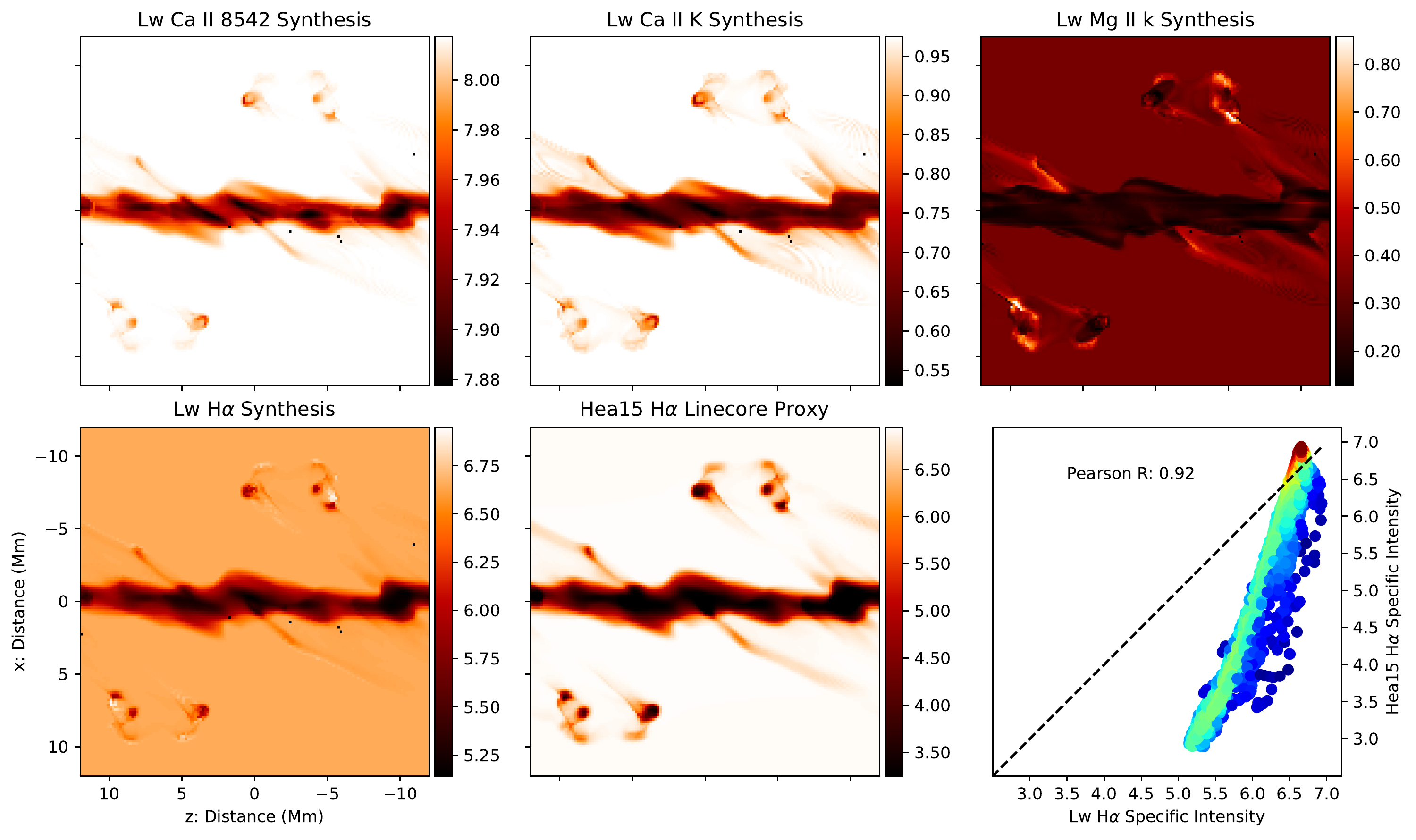}}
    \caption{The filament appearance of the \MpiAmr{} filament/prominence simulation according to the \Lw{} framework synthesis of various spectral lines. \textit{Bottom row}; Comparison between the Hydrogen H$\alpha$ line core synthesis and that of the proxy method of \citetalias{Heinzel:2015}; the pixel-by-pixel comparison in the rightmost panel, coloured according to the \ac{KDE} for the $\mathrm{log_{10}}(I_\nu)$ scatter, yields a Pearson R test score of 0.92. \textit{Top row}; The appearance of the simulated filament according to the \Caii{~8542~\&~K}, and \Mgii{~k} line cores respectively. All units $10^{-6}$~ergs~s$^{-1}$~cm$^{-2}$~sr$^{-1}$~Hz$^{-1}$.}
    \label{fig:filament_comparison}
\end{figure*}

As stated, the simulation was carried out on a uniform grid of 144$^3$ and so the extraction of a single vertical column is algorithmically trivial. 
Nevertheless, we intend to apply eventually this synthesis approach to more advanced simulations such as that presented in \citet{Jenkins:2022}, thus we have made use of the \ac{yt} framework which is generalised to consider \ac{AMR} arbitrarily. 
Herein, an `atmosphere' is extracted according to a user-defined ray (axis-aligned in this use case).
To construct these atmospheres, the \texttt{lightweaver.Atmosphere.make\_1d()} atmosphere constructor requires stratifications in height$^\dagger$, temperature$^\dagger$, microturbulent velocity, \ac{LOS} velocity$^\dagger$, electron number density, and total Hydrogen number density, where $\dagger$ indicates a property directly available within the \MpiAmr{} output. 
For the electron number density estimate, we use the \ac{NLTE} tables for $n_e$ provided in \citetalias{Heinzel:2015} - for total Hydrogen number density, we make use of the similar \ac{NLTE} ionisation degree $i$ tables also available from this study and compute according to $n_H = n_e / i$.
These tables reduce the dependency of the problem to the local temperature and pressure and so are perfectly suited here. 
Finally, the microturbulent velocity is set according to equations~13\,--\,16 of \citet{Heinzel:2001} with $\epsilon=0.5$.

We employ a 5 level + continuum Hydrogen atom with 10 bound-bound transitions, a 5 level + continuum Calcium~{\sc ii} atom with 5 bound-bound transitions, and a 10 level + continuum Magnesium~{\sc ii} atom with 15 bound-bound transitions \citep[the same used by][]{Leenaarts:2013a}.
Each of these atoms are derived from those distributed with RH\footnote{\url{https://github.com/han-uitenbroek/RH}} \citep{Uitenbroek2001}, and use the same atomic parameters as these models.

\section{Results} \label{s:Results}

\subsection{Synthesis} \label{ss:synthesis}

Following the setup described in Section~\ref{sss:two-part_geom}, each projection is constructed as a series of independent 1.5D atmospheres. 
For all syntheses, the \MpiAmr{} portion of the two-part atmospheres is inserted at a height of 5~Mm above the FAL-C atmosphere.
We therefore use the tabulated 10~Mm $n_\mathrm{e}$ values of \citet{Heinzel:2015} to initialise the equilibrium populations in approximately \ac{NLTE}.
Statistical equilibrium is then solved individually for each column from the simulation, before solving for the radiative transfer and emergent spectrum assuming an observer viewing parallel to the atmospheric stratification ($\mu_z=1$).
As detailed in Section~\ref{ss:lightweaver}, with these atomic models \Lw{} constructs the solar spectrum between 0\,--\,40000~\AA{}. 
One may then \textit{zoom} in on a portion of this wavelength range and inspect the appearance of any specific spectral line so long as the necessary transitions have been considered in statistical equilibrium. 
For this study, we will focus primarily on the line cores of Hydrogen~H$\alpha$, the Calcium~{\sc ii}~8542~\AA{} and H\&K, and the Magnesium~{\sc ii}~h\&k lines - hereafter referred to as \halpha{}, \Caii{~8542, H, K}, and \Mgii{~h, k}, respectively. 
Those atoms responsible for this selection of transitions are considered in \ac{NLTE}, in addition to a comprehensive \ac{LTE} background.
The initial collection of 144$^2$ fully independent atmospheres thus yields a 144~$\times$~144~$\times$~1617 spectral cube, where the number of wavelength points (1617) is computed from the wavelength quadrature specified by the atomic models, primarily defined to ensure that the integration of the radiative rates is correct.
Within \Lw{}, requesting a different set of `active' atoms will automatically compute the necessary sampling of wavelength points.

\begin{figure*}
    \centering
    \resizebox{1.0\hsize}{!}{\includegraphics{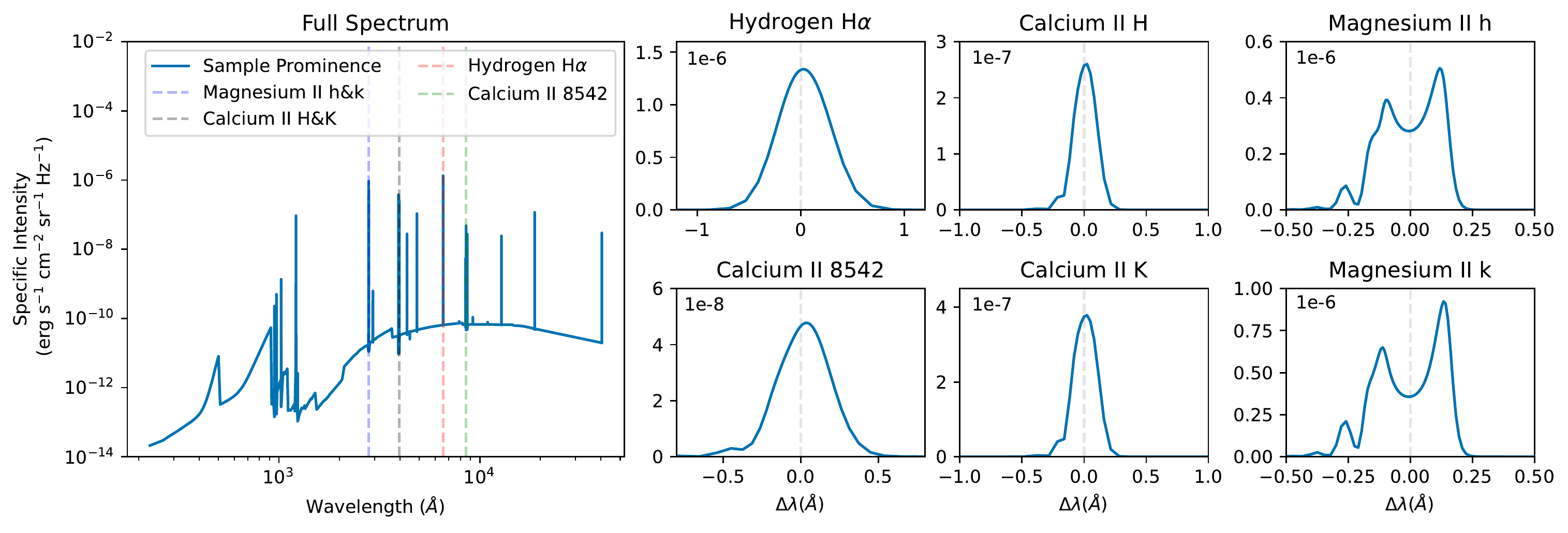}}
    \caption{Sample prominence synthesis for position (-10,7.5)~Mm within the `X projection' panel within the top row of Figure~\ref{fig:topology_example_filament_spectra}. On the right we show the sample of six lines typically used to observe prominences: \halpha{}, \Caii{~8542}, \Caii{~H\&K}, and \Mgii{~h\&k}.}
    \label{fig:example_prominence_spectra}
\end{figure*}

The solution procedure considers a total of 20,736 columns, taking $\approx33$ and $23$ hours of wallclock time (thus averaging five and four minutes per column) for the filament and prominence projection, respectively, on an Intel(R) Xeon(R) Silver 4210 CPU @ 2.20GHz 20 core/40 thread desktop.
In addition, we employ a cascade solution approach to maximise convergence across the projection \ac{FOV}: \ac{PRD} alone; \ac{CRD} with a second round of \ac{PRD} iterations; \ac{PRD} with collisional-radiative switching \citep[][]{Hummer:1988}; \ac{PRD} following a naïve spatial averaging of the neighboring eight columns.
If reached, none of the atmospheres converge under the spatial averaging approach.
Those atmospheres that either fail the cascade, or reach an iteration step greater than an arbitrary value of 740 (much higher than an average of 200) are rerun using an ad-hoc linear spatial upsampling of the primitive simulation variables to a resolution of 288 points in height. 
This equates to 42 and 101 atmospheres for the prominence and filament projections, respectively, from which only four and six atmospheres do not converge after upsampling, likely due to persistent insufficient sampling of the \ac{PCTR}.

\subsubsection{Filament Projection} \label{sss:filament}


In Figure~\ref{fig:topology_example_filament_spectra} we present a comparison between two profiles synthesised for stratifications characteristic of either a column containing a portion of the filament, or that of the FAL-C model.
As was the case for the isothermal test setup of Figure~\ref{fig:15d_vs_2part}, the example filament profile for each spectral window is clearly identifiable by the additional absorption compared to that of the \textit{chromospheric} synthesis. 
In general, for the filament formed using \MpiAmr{}, we find the \halpha{}, \Caii{} H\&K and \Mgii{} h\&k, to contain more significant absorption signatures \textit{i.e.}, larger contrast, than the \Caii{~8542}. 
Furthermore, the influence of the \ac{LOS} velocities on the synthesised profiles is clear, in particular for the \Caii{} and \Mgii{} resonance lines where the `horns' in each case are comparably asymmetric. We will not focus further on the shape and behaviour of the line wings in this study, a discussion for this is available within Section~\ref{s:Discussion}.

The bottom-left panel of Figure~\ref{fig:filament_comparison} presents the resulting 2D filament \halpha{} line core synthesis of the \MpiAmr{} simulation. 
The middle panel of the bottom row in the same figure presents the equivalent appearance of the simulation according to the \halpha{} line core proxy method of \citetalias{Heinzel:2015}.
Each method accounts for the \ac{LOS} projection of the local velocity during the \ac{RTE} integration; for the \citetalias{Heinzel:2015} method this is restricted to a decrease in the line core opacity for strong flows. 
The bottom-right panel provides a 1-1 comparison between these \Lw{} and \citetalias{Heinzel:2015} \halpha{} syntheses, the Pearson R score indicating a strong positive correlation.
The discrepancy between the near-linear \ac{KDE} and a 1-1 relation is explored in Section~\ref{s:Discussion}.
The middle row of Figure~\ref{fig:filament_comparison} then presents the \Lw{}-synthesised appearance of the \MpiAmr{} simulation according to the \Caii{~8542~\&~K}, and \Mgii{~k} line cores; the longer wavelength counterparts of the resonance lines appear almost identical and so are omitted (cf. Figures~\ref{fig:filament_spectrum_example}\&\ref{fig:example_prominence_spectra}). 

\begin{figure*}
    \centering
    \resizebox{1.\hsize}{!}{\includegraphics{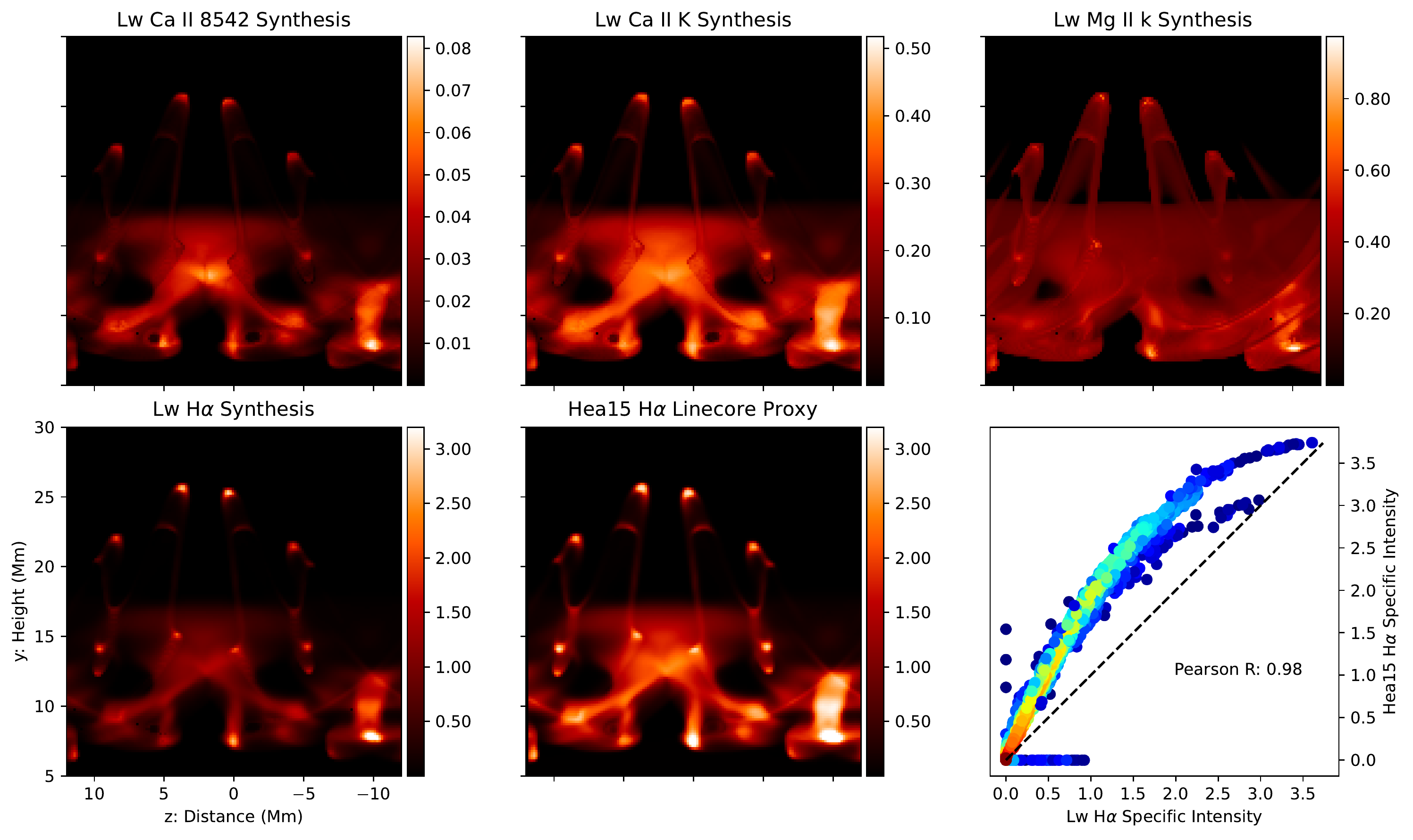}}
    \caption{The prominence appearance of the \MpiAmr{} filament/prominence simulation according to the \Lw{} framework synthesis of various spectra lines. \textit{Bottom row}; Comparison between the Hydrogen H$\alpha$ line core synthesis and that of the proxy method of \citetalias{Heinzel:2015}; the pixel-by-pixel comparison in the rightmost panel, coloured according to the \ac{KDE} for the $\mathrm{log_{10}}(I_\nu)$ scatter, yields a Pearson R test score of 0.98. \textit{Top row}; The appearance of the simulated filament according to the \Caii{~8542~\&~K}, and \Mgii{~k} line cores respectively. All units $10^{-6}$~ergs~s$^{-1}$~cm$^{-2}$~sr$^{-1}$~Hz$^{-1}$.}
    \label{fig:prominence_comparison}
\end{figure*}

\subsubsection{Prominence Projection} \label{sss:prominence}

For the prominence projection, we consider the \MpiAmr{} simulation to be positioned exactly at the solar limb. 
The corresponding \Lw{} atmosphere thus describes a horizontal cut through the simulation. 
For a prominence projection, the \textit{lower} boundary condition now refers to radiation incident on the \textit{backside} of the prominence, and the frontside equivalently for the upper boundary.
Geometrically, the boundary conditions are constructed following a change in coordinate system, since $\mu_z = 1$ now considers a ray that runs parallel, rather than perpendicular, to the solar surface, as shown in Figure~\ref{fig:problem_geometry}.
Finally, we adopt a zero incident radiation assumption for those $\mu$ angles representing a ray that intersects the theoretical corona. 
An example spectrum emergent from the prominence atmosphere is presented in Figure~\ref{fig:example_prominence_spectra}.

As detailed, the 1.5D approximation for the prominence atmospheres considers radiation incident on the stratification from both infront (the top) and behind (bottom) but contains no information about the radiation incident from directly below or indeed any adjacent atmosphere between it and the solar surface.
The influence of this approximation, and associated limitations imposed on any conclusions, will be discussed in Section~\ref{s:Discussion}. 
In the same arrangement as the filament syntheses of Figure~\ref{fig:filament_comparison}, Figure~\ref{fig:prominence_comparison} presents the corresponding prominence syntheses. 
Once again, the Pearson R score finds a strong positive correlation between the \Lw{} and \citetalias{Heinzel:2015} \halpha{} syntheses.
Zero values for \Lw{} \halpha{} indicate those pixels that did not converge; zero values for \citetalias{Heinzel:2015} \halpha{} demonstrate the lack of opacity donated from the $n_2$ tables of \citetalias{Heinzel:2015} so as to produce an emission signature.


\begin{figure*}
    \centering
    \resizebox{0.92\hsize}{!}{\includegraphics{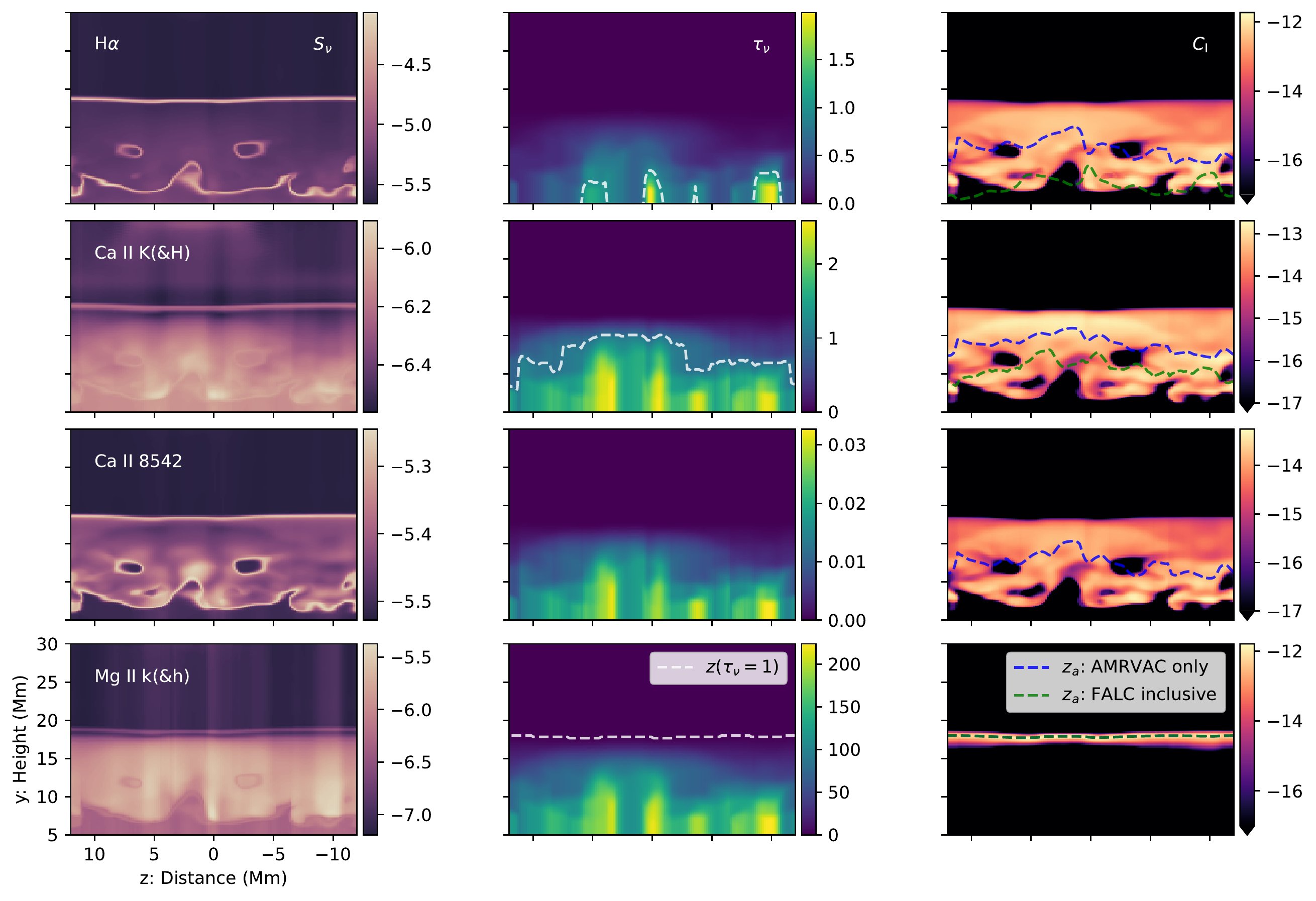}}
    \caption{Line core formation properties within a cut through $x=0$~Mm of the filament synthesis. The source function $S_\nu$ and integrated opacity $\tau_\nu$, and associated contribution function $C_\mathrm{I}$ are shown for four of the six line core transitions shown in Figures~\ref{fig:filament_comparison}~\&~\ref{fig:prominence_comparison}. The dashed lines overlaid on the $\tau_\nu$ and $C_\mathrm{I}$ quantities are the location of $\tau_\nu=1$, and the average formation height $z_\mathrm{a}$ quantity calculated for both an \MpiAmr{}-only and FAL-C-inclusive stratification, respectively, and overlap in the case of \Mgii{~h\&k}. $S_\nu$ and $C_\mathrm{I}$ quantities in units of $\mathrm{log}_{10}$(ergs~s$^{-1}$~cm$^{-2}$~sr$^{-1}$~Hz$^{-1}$).}
    \label{fig:formation_properties_filament}
\end{figure*}

\subsection{The 1.5D formation of various spectral lines}\label{ss:1.5d_formation}

\subsubsection{Filament Projection}\label{sss:1.5d_formation_filament}

In Section~\ref{sss:15d_two-part_comparison} we briefly highlighted a qualitative agreement in Figure~\ref{fig:15d_aniso_temperature} between the source function of our two-part, isothermal \& isobaric model and that of \citet{Heinzel:1995}. 
However, such an idealised stratification is absent from those 1.5D atmospheres drawn from the nonlinear \MpiAmr{} simulation.
Hence, the resulting stratification of parameters key to the emergent line core intensities of the various spectral lines presented in Figures~\ref{fig:filament_comparison}~\&~\ref{fig:prominence_comparison} must necessarily differ.
Figure~\ref{fig:formation_properties_filament} explores the spatial variation of formation properties for the line cores of these transitions within a cut that runs vertically along $x=0$~Mm, a location approximately representative of the filament spine. 
Although the incorporated atmospheres are generally highly structured, their distance from the lateral edge of the filament, and hence significant additional incoming radiation, suggests the 1.5D approximation would have the least (but of course nonzero) influence here (we will discuss this in more detail in Section~\ref{s:Discussion}).
In the left-hand column of Figure~\ref{fig:formation_properties_filament} we detail the frequency $\nu$ and direction $\mu$ dependent local source function, calculated as given in \citet{Osborne:2021},
\begin{equation}
    S_{\nu,\mu}\equiv \frac{\eta^\mathrm{tot}_{\nu,\mu} + \sigma_\nu J_\nu}{\chi^\mathrm{tot}_{\nu,\mu}}, \label{eq:source_function}
\end{equation}
where $\eta^\mathrm{tot}$ and $\chi^\mathrm{tot}$ are the total (summation over all species) emissivity and opacity, respectively, $J_\nu$ the angle-averaged intensity, and $\sigma_\nu$ the continuum scattering coefficient. 

The source functions for the \halpha{} and \Caii{~8542} lines describe an enhancement at the upper and lower edges of the filament, where the associated sharp gradients in temperature represent the \ac{PCTR} and thus an enhancement in the $n=2$ levels of Hydrogen in these locations \citep[cf.][]{Anzer:1999}. 
Within the filament, however, the source function is comparatively weak; for the case of \Caii{~8542} the source increases only in the locations of a temperature minimum of order 10$^3$~K. 
The (H\&)K line of \Caii{} contains clear enhancements in the upper and lower \acp{PCTR} but with the lower characterised by a broader, more gradual drop of magnitude with height.
The source functions for the \Mgii{~(h\&)k} line has its peaks, instead, embedded with a broad distribution throughout the body of the filament. In general, the longer wavelength components of the resonance lines share near-identical distributions of each source parameter with their counterparts and are once more omitted here.

A measure of the integrated opacity $\tau_\nu$ (along each filament column in the negative $y$ direction) for the selection of spectral lines (at the line core) is shown in the second column of Figure~\ref{fig:formation_properties_filament}. 
For \halpha{}, we find a significant portion of the spine to contain $\tau_\nu \lessapprox 1$; for \Caii{~8542} we find the opacity to be extremely low and is key to the low contrast seen in Figure~\ref{fig:filament_comparison}; for the \Caii{~H} line we find a similar range as for \halpha{} and a mildly optically-thick profile for \Caii{~K}; the \Mgii{~h\&k} lines then exhibit their characteristic optically-thick properties with a total opacity several orders of magnitude above the others considered here.


A comparison between the corresponding contribution functions is shown in the right column of Figure~\ref{fig:formation_properties_filament}. This quantity is essentially a modified form of the integrand in Eq.~\ref{eq:formal_RTE} that represents the contribution of a local volume (voxel at position $z'$) to the emergent specific intensity $I_\nu(z_\mathrm{t})$ (measured by an observer at position $z_\mathrm{t}$, the top of the \MpiAmr{} simulation domain). Calculated as,
\begin{equation}
    C_\mathrm{I}(\nu,z') \equiv \left.\frac{d I_\nu(z_\mathrm{t})}{dz}\right|_{z=z'} = \frac{\chi^\mathrm{tot}_\nu(z')}{\tau_\nu(z')} \cdot S_\nu(z') \cdot \tau_\nu(z')e^{-\tau_\nu(z')}, \label{eq:contribution_function}
\end{equation}
for which we have assumed $\tau_\nu(z_\mathrm{t})=0$, and the \ac{LOS} is taken to be parallel to the filament atmospheres \textit{i.e.}, exactly vertical. Indeed, for both the filament or prominence synthesis all quantities are calculated such that $\mu=1$ and so the $\mu$ index will be hereafter dropped for brevity. 
Here we see clearly how misleading conclusions may be, if drawn from the source function alone, since it fails to convey how a local peak in $S_\nu$ propagates through the remainder of the atmosphere towards the observer \citep[cf. the second component of Eq.~\ref{eq:formal_RTE}, and][]{Carlsson:1997}. 
For \halpha{}\,--\,\Caii{~8542}, we find similar distributions of the contribution function spread throughout the individual filament columns as a consequence of the comparable $\tau_\nu$ ranges; the \Caii{~8542} contribution function does however peak several orders of magnitude lower. 
The significantly higher optical thickness of \Mgii{~h\&k} leads to a contribution function that is heavily peaked in the upper \ac{PCTR}, despite the broad source function, with very limited contribution from the internal layers of the filament.

\begin{figure}
    \centering
    \resizebox{\hsize}{!}{\includegraphics{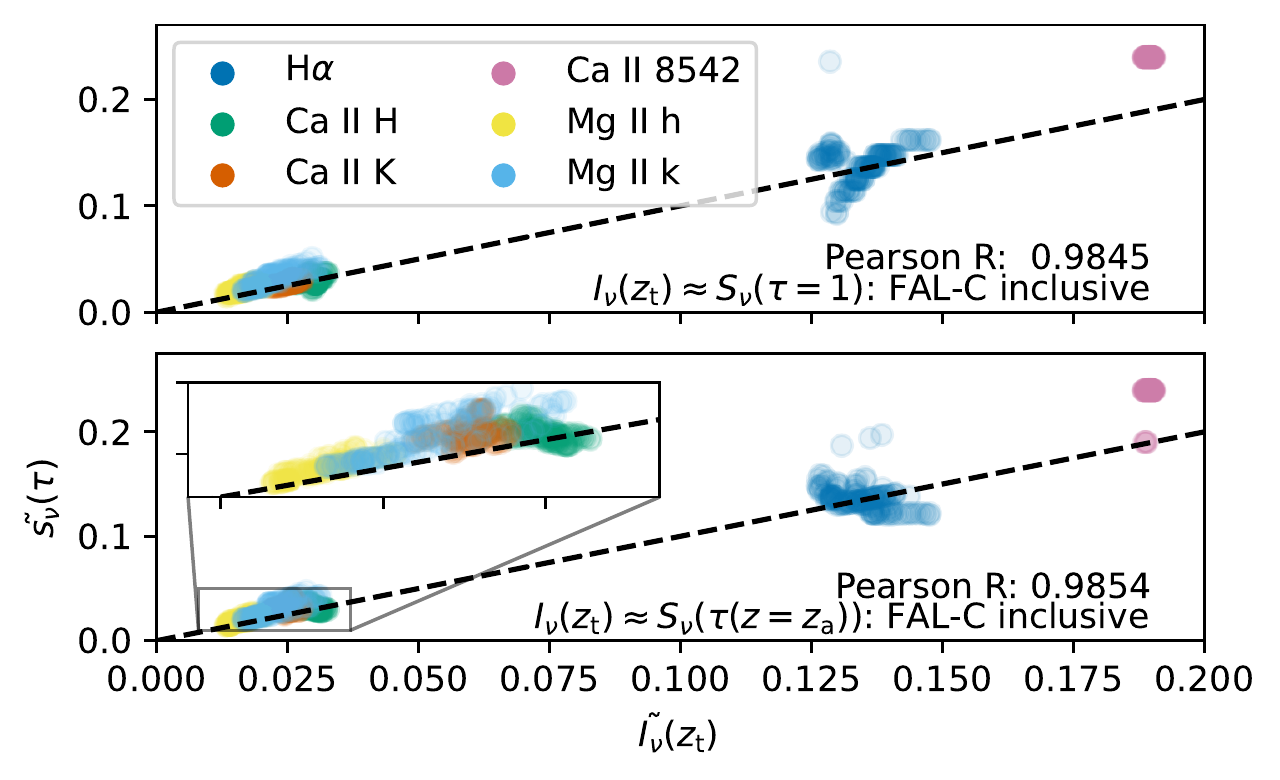}}
    \caption{Relationship between emergent intensity and different assumptions for where the majority of the emergent intensity comes from within the spatial distribution of the source function. The dashed-black line indicates in each case a 1\,--\,1 relation between $\tilde{I}$ and $\tilde{S}$, wherein the $\sim$ signifies the quantity is normalised by a nearby continuum value.}
    \label{fig:emergent_source_assumptions}
\end{figure}

When considering from where the \textit{majority} of information encoded within a spectral line is sourced, one often finds reference to the \ac{EB} approximation ($I_\nu(z_\mathrm{t}) \approx S_\nu(\tau_\nu = 1)$) along the \ac{LOS} ($\mu=1$); for an optically-thick medium, the emergent intensity is approximately sourced from a location one photon mean-free path from the observer \citep[see Figure~36 of][]{Vernazza:1981}. 
The top panel of Figure~\ref{fig:emergent_source_assumptions} tests the \ac{EB} approximation for the filament atmospheres of Figure~\ref{fig:formation_properties_filament}, merged with the FAL-C atmosphere. 
Previously, \citet{Leenaarts:2012a} remarked that there exists no perfect measure for the formation height of a given line within the chromosphere on account of the typically broad contribution functions. From Figure~\ref{fig:formation_properties_filament} we find this comment similarly relevant here within filaments. 
Instead, for quiet-Sun chromospheric modelling, these authors marginally favoured the \textit{average formation height} quantity given by,
\begin{equation}
z_{\mathrm{a}}=\frac{\int_{z_{\mathrm{b}}}^{z_{\mathrm{t}}} z^{\prime} \chi^\mathrm{tot}_\nu S_\nu \mathrm{e}^{-\tau_\nu} d z^{\prime}}{\int_{z_{\mathrm{b}}}^{z_{\mathrm{t}}} \chi^\mathrm{tot}_\nu S_\nu \mathrm{e}^{-\tau_\nu} d z^{\prime}},
\end{equation}
\textit{i.e.}, the average height weighted by the contribution function of Eq.~\ref{eq:contribution_function}. The middle and bottom panels of Figure~\ref{fig:emergent_source_assumptions} then compare the same quantities as the top panel of the same figure, but with $S_\nu$ sampled instead at height $z_\mathrm{a}$ calculated for atmospheres both excluding and including the contribution of the FAL-C chromosphere. 
From the Pearson R test scores, we conclude similarly that the \ac{EB} and $z_\mathrm{a}$ approximations fare equally - both struggling with the comparably optically-thin \halpha{} and \Caii{~8542} line cores - but with the latter performing marginally better overall.

\begin{figure}
    \centering
    \resizebox{0.95\hsize}{!}{\includegraphics{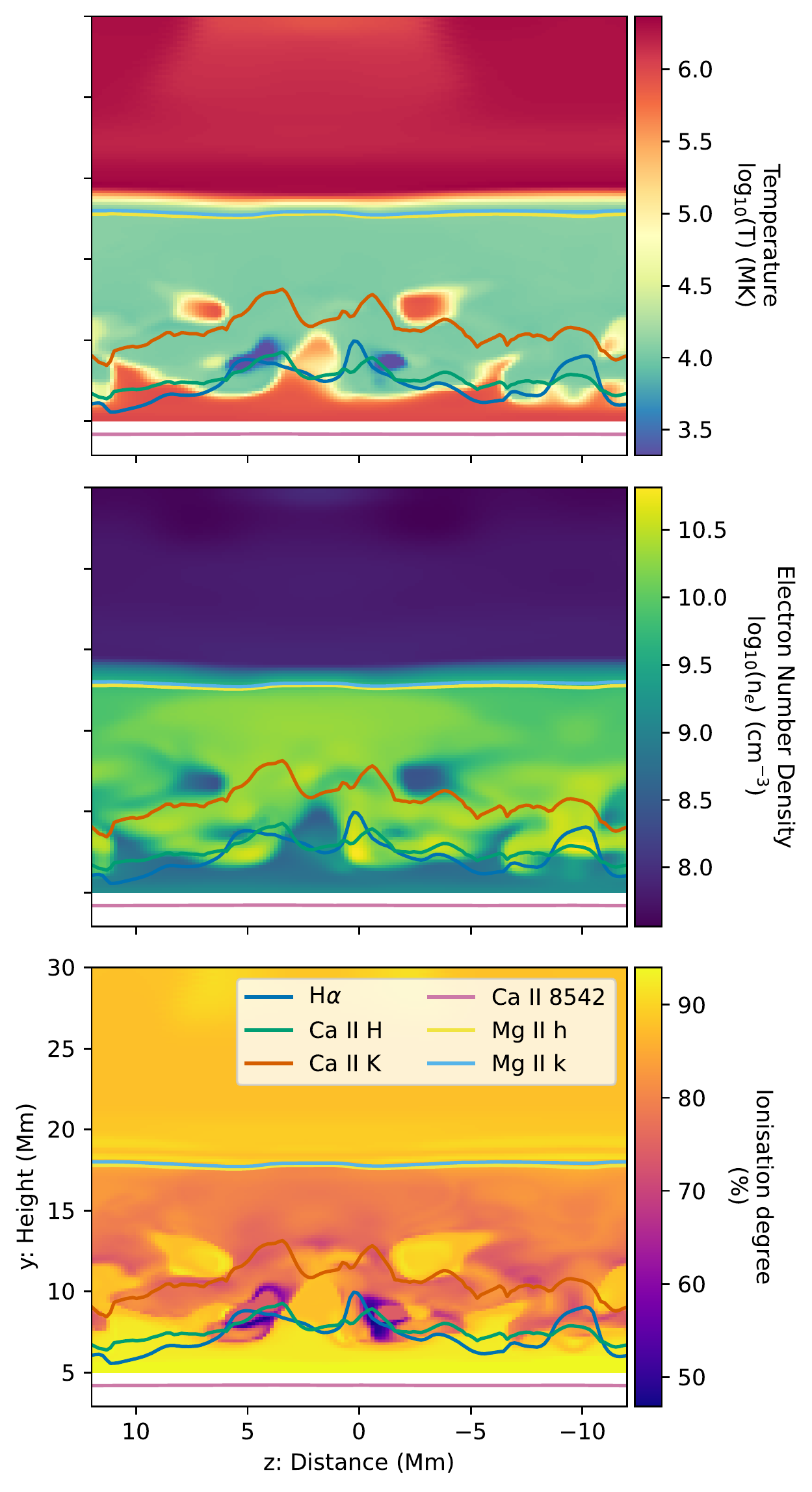}}
    \caption{The line core formation of \halpha{}, \Caii{~8542, H\&K}, and \Mgii{~h\&k} within the \MpiAmr{}-simulated filament. The \textit{average formation height} $z_\mathrm{a}$ calculated inclusive of the FAL-C chromospheric model for the line cores as previously shown in Figure~\ref{fig:formation_properties_filament}, overplotted on co-spatial distributions of the primitive quantity of temperature, and \citetalias{Heinzel:2015}-inferred electron number density and ionisation degree.}
    \label{fig:average_formation_properties_filament}
\end{figure}

Comparing the column height of $z_\mathrm{a}$, we show in the rightmost column of Figure~\ref{fig:formation_properties_filament} that the \MpiAmr{}-only $z_\mathrm{a}$ traces approximately the geometrical middle of the regions of large contribution function. 
Peaks in the source function within the FAL-C atmosphere, as noted in Figure~\ref{fig:15d_vs_2part}, then lead to a notable decrease in $z_\mathrm{a}$. 
Nevertheless, the approximation of the average formation height remains clearly influenced by the presence of the filament within the atmosphere for all lines except \Caii{~8542}. 
In Figure~\ref{fig:average_formation_properties_filament}, we overplot $z_\mathrm{a}$ inclusive of the FAL-C model for the filament synthesis on co-spatial cuts of the temperature, and (\citetalias{Heinzel:2015} inferred) electron density and ionisation degree within the \MpiAmr{} simulation. 
Indeed here, we find $z_\mathrm{a}$ for \Caii{~8542} to lie below the \MpiAmr{} simulation domain, between it and the FAL-C chromosphere, with a narrow range of 4.19\,--\,4.23~Mm and shape nearly identical to its \Caii{~H\&K} counterparts.
In this way, the \Caii{~8542} line remains only weakly influenced by the filament atmosphere, in accordance with its narrow contrast found in Figure~\ref{fig:filament_comparison} \citep[cf. chromospheric formation height][]{Uitenbroek:1989a,Leenaarts:2009a,Diazbaso:2019}. 
Based on the assumptions laid out above, we thus find that the line cores of the selected spectral lines would form in this simulated filament, with increasing $z_\mathrm{a}$, as \Caii{~8542} $<$ \halpha{} $\approx$ \Caii{~H} $<$ \Caii{~K} $<$ \Mgii{~h\&k}; the identical ordering, ignoring the highly varying small scale behaviour, can also be reached from the $z(\tau_\nu=1)$ approximation (cf. $\tau_\nu$ panel of Figure~\ref{fig:formation_properties_filament}).

\begin{figure*}
    \centering
    \resizebox{1.\hsize}{!}{\includegraphics{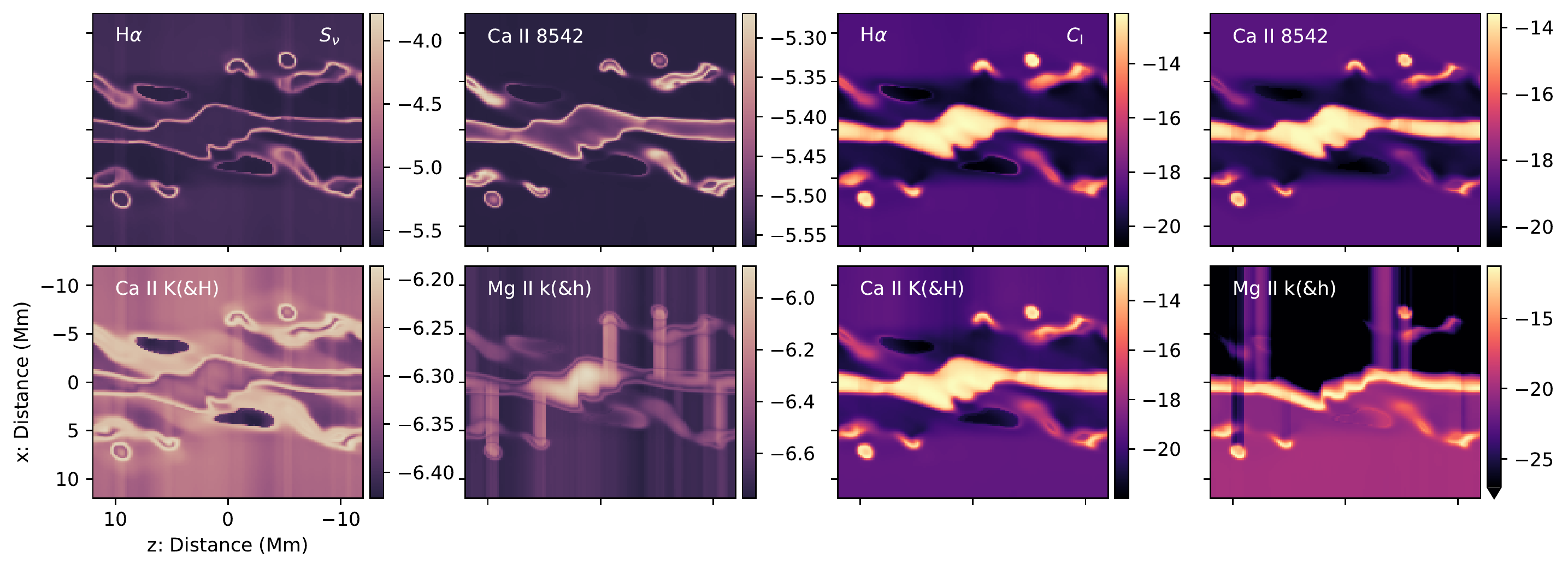}}
    \caption{The spatial distribution of line core formation properties for the prominence synthesis in a cut through a height of $y\approx13$~Mm. The left four panels detail the source function $S_\nu$ and the right four the equivalent contribution function $C_\mathrm{I}$, the latter calculated for a \ac{LOS} aligned with the negative-$x$ direction. All quantities in units of $\mathrm{log}_{10}$(ergs~s$^{-1}$~cm$^{-2}$~sr$^{-1}$~Hz$^{-1}$).}
    \label{fig:formation_properties_prominence}
\end{figure*}

\subsubsection{Prominence Projection}\label{sss:1.5d_formation_prominence}

Quantifying the formation height, or rather \textit{depth}, as in Section~\ref{sss:1.5d_formation_filament}, for a single 1.5D prominence column is comparably trivial on account of the lack of background illumination. 
Hence, for an entirely optically-thin, isothermal/isobaric prominence atmosphere, the contribution function peaks at its geometrical centre. 
As already indicated in Figure~\ref{fig:formation_properties_filament}, the average formation height for an increasingly optically-thick atmosphere will then be equivalently skewed towards those layers that are closest to the observer; for an arbitrary observer this equates similarly to a symmetric contribution function about the middle of the prominence atmosphere \citep[cf.][]{Heinzel:2005, Gunar:2007a}. 
For the prominence that is presented within the \ac{MHD} simulation, there are a wide range of optically-thin\,--\,optically-thick profiles as would be expected within an observed solar prominence.
Furthermore, and as demonstrated by \citet{Gunar:2008} for the very optically-thick Lyman-$\alpha$ line, including the \ac{PCTR} contribution from an ensemble of threads each with their own \ac{LOS} velocity contribution is of paramount importance to recover the asymmetric nature of such an optically-thick spectral line. 
The discrete condensations of finite extent shown in Figure~\ref{fig:topology_example_filament_spectra} mean that this multi-threaded property is also an important feature of the \ac{MHD} model presented here.

\begin{figure}
    \centering
    \resizebox{1.\hsize}{!}{\includegraphics{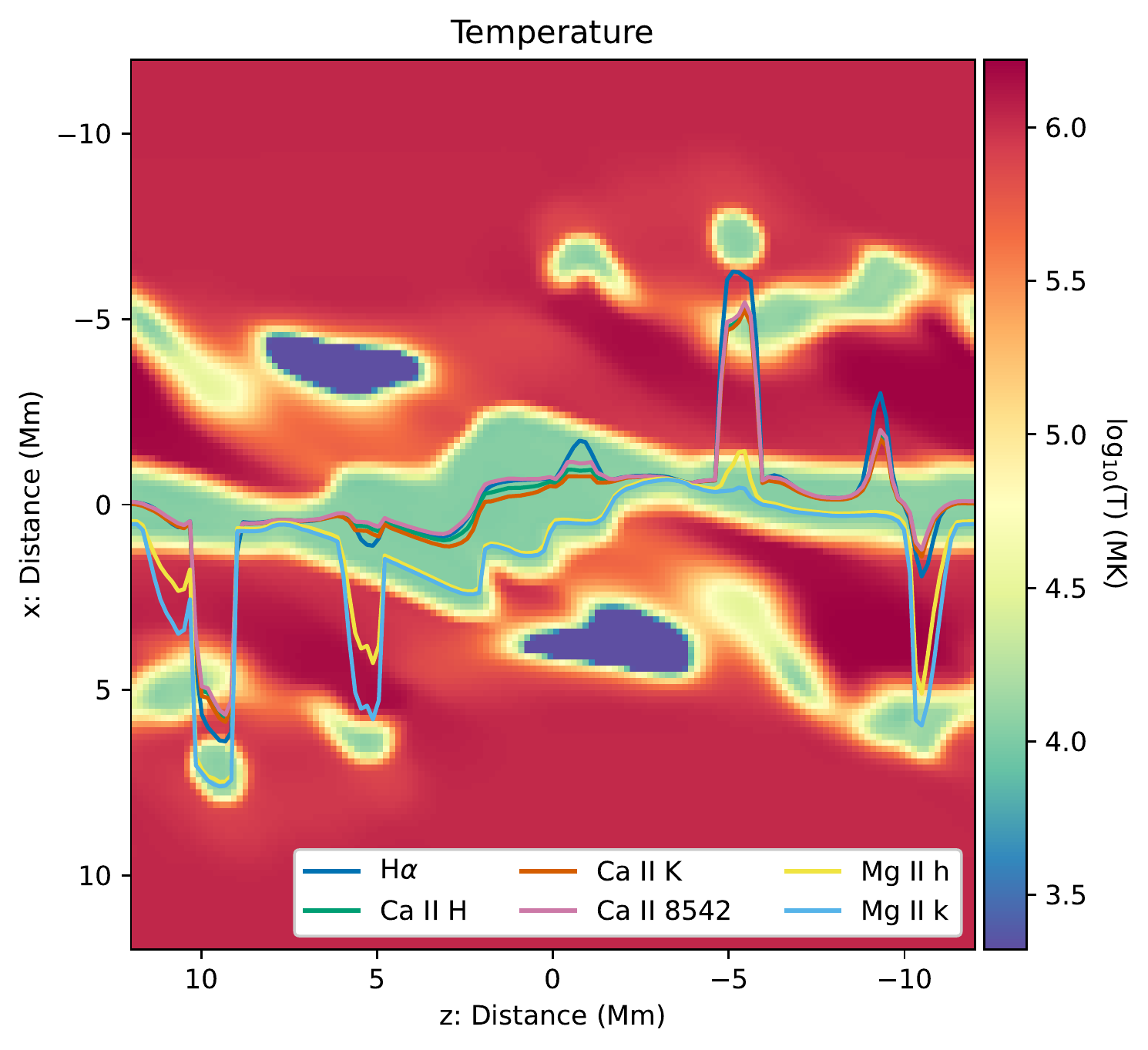}}
    \caption{The line core formation of \halpha{}, \Caii{~8542, H\&K}, and \Mgii{~h\&k} within the \MpiAmr{}-simulated prominence. The \textit{average formation depth} $z_\mathrm{a}$ of the line cores in the same horizontal cut as Figure~\ref{fig:formation_properties_prominence}, overlaid on the primitive temperature quantity.}
    \label{fig:average_formation_properties_prominence}
\end{figure}

The two sets of six panels in Figure~\ref{fig:formation_properties_prominence} detail the source $S_\nu$ and contribution $C_\mathrm{I}$ functions, as in Figure~\ref{fig:formation_properties_filament}, within a horizontal cut through the prominence synthesis at a height of $\approx$~13~Mm (cf. Figure~\ref{fig:prominence_comparison}). 
Herein we find the source function to be highly structured for each line core. 
To the best of our knowledge, and albeit perhaps unsurprising, this is the first time such highly structured source functions have been displayed for solar prominences/filaments \citep[with the only other study showing a similar degree of variability in space being that of][for the integrated intensity]{Labrosse:2016}.
For the comparably optically-thin \halpha{}\,--\,\Caii{~8542} lines, the largest $S_\nu$ values, without exception, are located within the hotter, less-dense \ac{PCTR} regions. 
A comparison with their contribution functions (calculated for an observer at $x=12$~Mm) yields limited overlap; very few locations with an enhanced source function dominate the emergent intensity in this 1.5D approximation. 
Instead, only those \ac{PCTR} enhancements located immediately adjacent to a strong density gradient have a non-negligible, but nevertheless weak, contribution donated by $\tau_\nu$ in $C_\mathrm{I}$. 
Elsewhere within the regions that contain an elevated contribution function, $S_\nu$ remains relatively constant for a given column.
For the \Mgii{~h\&k} lines, peaks in the contribution functions overlap exactly with the sharp gradients in the source function at the \ac{PCTR} boundaries. 
A lack of elevated specific intensity within the core of the prominence in Figure~\ref{fig:prominence_comparison} demonstrates how the large value of $\tau_\nu$ for \Mgii{~h\&k} then prevents the \ac{LOS} emergent intensity from being directly encoded with the enhanced $S_\nu$ values within the core of the prominence.

Overlaid on the temperature structure of the \ac{MHD} simulation, Figure~\ref{fig:average_formation_properties_prominence} presents the average formation depth of \halpha{}, \Caii{~8542, H\&K}, and \Mgii{~h\&k} for the same cut as Figure~\ref{fig:formation_properties_prominence}. 
At this height within the simulation, the aforementioned multi-threaded components along the \ac{LOS} are well-captured. 
As for the filament synthesis, the depth to which a \ac{LOS} reaches within the prominence atmosphere depends on the chosen line core. 
For all columns in this slice, the h\&k line cores of \Mgii{} form at the shallowest layers, followed by the equivalent resonance lines for \Caii{}, and finally \Caii{~8542}. 
The \textit{location} of the \halpha{} line core, on the other hand, varies significantly for those columns that contain multiple threads, influencing it to be positioned at either shallower or deeper layers than the three \Caii{} line cores.
As there exists many columns within the prominence synthesis where some of the line cores do not encounter $\tau_\nu=1$ conditions, no equivalent estimation is possible with the \ac{EB} approximation.

\section{Discussion} \label{s:Discussion}

This manuscript presents our recent efforts in applying the \Lw{} \ac{NLTE} synthesis framework to the modelling of prominences or filaments, including correct handling of the geometry and necessary irradiated boundary conditions. We detailed the 1.5D plane-parallel approximation that, although geometrically trivial, represents a necessary first step towards a full 3D synthesis \citep[akin to][]{Leenaarts:2012a,Bjorgen:2019}. We began by showing that the standard boundary conditions for simplified, plane-parallel atmospheres implemented within the \Lw{} synthesis tool are unsuitable for a stratification including an elevated prominence/filament atmosphere. 

For the filament case, the horizontal invariance of the 1.5D geometry led to a significant portion of the radiation emitted below the filament and within the chromosphere being absorbed and in turn scattered back towards lower heights, resulting in the artificial `radiation pumping' of these intermediate layers. Such an undesirable, unphysical feature was previously hinted at by \citet{Paletou:1993}. This was subsequently overcome by pre-synthesising an assumed ThermalisedRadiation + FAL-C chromosphere + ZeroRadiation model, and feeding it into the filament columns as a lower boundary condition, whilst simultaneously taking into account limb-darkening from this model. 
To ensure a smooth limb-darkening function, we adopted a cone-averaged intensity representation for each ray-set in the base quadrature as in Figure~\ref{fig:fil_prom_BCs}. 
For the filament projection we assumed a cylindrically-symmetric distribution for the angularly-dependent background illumination. 
For the prominence projection, on the other hand, the boundary conditions necessarily considered a change in orientation, such that radiation was input along $\mu_x$ rather than $\mu_z$ (albeit maintaining the $\mu_z$ definitions for the series of sampling cones), removing the cylindrical symmetry. 

First and foremost, the results presented in Figures~\ref{fig:filament_comparison}~\&~\ref{fig:prominence_comparison} demonstrate that the 1.5D \Lw{} framework with modified boundary conditions is capable of synthesising a \MpiAmr{} coronal simulation including a flux rope + filament/prominence system in a variety of spectral lines. 
The contrast appearance of the simulated filament and prominence are consistent with observations; a comparison between our \halpha{} and \Caii{~H} prominence appearance and equivalent observations of \citet{Gunar:2014} is particularly satisfying. The correlation plots for each \halpha{} projection then demonstrate a clear order of magnitude agreement between \Lw{} and the 1.5D model of \citetalias{Heinzel:2015}.

\subsection{Comparison against \halpha{} proxy method of \citetalias{Heinzel:2015}}

In both \citet{Jenkins:2021, Jenkins:2022}, the authors employed the method of \citetalias{Heinzel:2015} to construct specific intensity representations of their simulations. This method takes advantage of a series of pre-computed tables for the empirical relationship between electron number density squared $n_e^2$ and the population density of the second level of the Hydrogen atom $n_2$. Denoted as $f$, \citetalias{Heinzel:2015} provide the variation of this scaling for a range of discrete heights throughout the solar atmosphere, in addition to the equivalent tables for ionisation degree $i$. The $f$ and $i$ quantities then rely simply on the local temperature and pressure, reducing the problem to a simple lookup operation for all grid positions before an arbitrary \ac{LOS} integration. The robustness of this approach has already been demonstrated in multiple previous studies, and again here, to yield smooth variations in intensity across the synthesised \ac{FOV} \citep[][]{Gunar:2015, Claes:2020, Zhou:2020, Jenkins:2021, Martinez-Gomez:2022}.

The \Lw{} synthesis of the filament and prominence presented in Figures~\ref{fig:filament_comparison}~\&~\ref{fig:prominence_comparison}, despite being collections of independent 1.5D atmospheres, produces smooth variations in intensity across the \ac{FOV}. 
This highlights how the two-part model constructed and implemented within \Lw{} is not only suitable for dealing with solar (stellar) atmosphere stratifications involving an elevated filament/prominence, but also the subtle variations therein. 
For the filament projection specifically, the smaller, globular structures located at higher elevations, see ($10>x>3$ , $8>y>6$)~Mm, are also recorded within the \Lw{} synthesis. 
The positive contrast recorded here is a consequence of accounting for the interplay between local velocity, temperature, and density and its influence on the line core opacity. Such a property cannot be approximated with the method of \citetalias{Heinzel:2015} as the influence of \ac{LOS} velocities is restricted to a decrease in line core opacity up to the value of the assumed background intensity.
Elsewhere, for both the filament and prominence projections, the relative intensity variations present within the \Lw{} \halpha{} synthesis are similar to those of the \citetalias{Heinzel:2015} approximate method.
In the accompanying correlation plots, we do, however, find a clear increase/decrease in intensity throughout the filament/prominence bodies, respectively.

Outside of the filament, the intensity of the line core differs between the methods as a consequence of the assumed background illumination. 
Following \citetalias{Heinzel:2015}, the chromospheric background adopts a fixed line core intensity following the disk-centre average measurements by \citet{David:1961}. 
For the filament two-part model in \Lw{}, we employed instead the fully stratified, and height/angle-dependent FAL-C model which is shown in Figure~\ref{fig:topology_example_filament_spectra} to yield a slightly deeper line core intensity. 
For the background intensities within the prominence projections, a near-exact agreement is found between the methods due to the zero background illumination and optically-thin properties along the \ac{LOS}. Those locations shown in the bottom-right panel of Figure~\ref{fig:prominence_comparison} to not adhere to the 1-1 trend have already been attributed to a lack of opacity donated from the $n_2$ tables of \citetalias{Heinzel:2015}.

To address the difference in line core intensity between the two methods within the filament/prominence bodies, recall that the approximate synthesis method of \citetalias{Heinzel:2015} considers a single, constant value for the source function along the \ac{LOS}, see Figure~\ref{fig:15d_vs_2part}. 
This assumed value then varies only in height \textit{i.e.}, constant or varying from column to column for the filament and prominence projections, respectively. 
Furthermore, the synthesis method of \citetalias{Heinzel:2015} uses the Lambertian (non-limb-darkened) approximation for the \textit{dilution factor}, additionally reported in \citet{Jejcic:2009} to be slightly \textit{too large}. 
Applying a corrective, ad-hoc fractional multiplication factor to the source quantity leads only to a systematic, linear shift in the resulting intensities. 
For the filament synthesis this leads to a darker filament and an equivalently darker prominence, simultaneously better-aligning the \Lw{} and \citetalias{Heinzel:2015} prominence syntheses whilst increasing the discrepancy between the filament syntheses. 
Hence, whilst the differing assumptions for $S_\nu$ remain an important consideration, in particular for the prominence synthesis, it cannot solely explain the mismatch. 

The scatter plots of Figures~\ref{fig:filament_comparison}~\&~\ref{fig:prominence_comparison} detail the offset between the \Lw{} and \citetalias{Heinzel:2015} syntheses to be neither linear nor systematic across the \ac{FOV} of the two projections. 
The amplitude of the absorption coefficient (equation 4 of \citetalias{Heinzel:2015}) defines, to first order, the amplitude of the line core absorption/emission properties for filaments/prominences synthesised using their method. 
Since this term depends on a normalised Gaussian and the tabulated second level population $n_2$ of Hydrogen, only the latter is capable of influencing the magnitude carried forward into the integration.
Furthermore, a linear change in $n_2$ leads to the aforementioned bifurcated influence on the final synthesis; a decrease in $n_2$ corresponds to an increase/decrease in the filament/prominence specific intensity, respectively. 
It is from the original correlations of \citet{Heinzel:1994} that \citetalias{Heinzel:2015} tabulated this $f$ ratio. 
However, the converged solutions found through the \Lw{} synthesis indicate these derived $n_2$ levels to be between $\approx$~35\,--\,55\% of the tabulated values from \citetalias{Heinzel:2015}. 
Furthermore, this fractional difference is non-constant and highly structured in height. 
We thus find this fractional difference in the fitted function $f$ to be the first-order cause for the discrepancy between the methods as shown in Figures~\ref{fig:filament_comparison}~\&~\ref{fig:prominence_comparison}.

\citet{Heinzel:2015} state explicitly the applicability of their approximate method to only those prominences/filaments that are approximately optically-thin $\tau_\nu\leq1$. 
For \halpha{}, we find this to be a reasonable conclusion; by comparing to our \Lw{} \ac{NLTE} synthesis we have been able to demonstrate that such an approximate method succeeds well even for those columns within the simulated filament/prominence that reach $\tau_\nu \gtrapprox 1$. However, such a statement is thusfar valid only for the 1.5D geometry.
Moreover, Figure~\ref{fig:filament_comparison} demonstrates such approximate methods will surely struggle to represent the full 1.5D \ac{NLTE} solution since they do not consider the intricate variations of velocities or $S_\nu$, and their accumulation $C_\mathrm{I}$, throughout a column. 
Indeed, as $\tau_\nu$ increases yet further for the \Caii{~H\&K} and \Mgii{~h\&k} examples, whether we consider the filament or prominence projections, their profiles presented in Figures~\ref{fig:topology_example_filament_spectra}~\&~\ref{fig:example_prominence_spectra} can be highly non Gaussian on account of this large $\tau_\nu$, $v_\mathrm{\ac{LOS}}$ encoding, and the subsequent non-local effects on $S_\nu$ along even a single column. 
An extension to higher dimensionality will surely demonstrate further how such approximate synthesis methods remain suitable only for sufficiently low-$\tau_\nu$ filaments/prominences, and will likely never be realisable for lines such as \Mgii{~h\&k}.

\subsection{1.5D line formation within solar filaments and prominences}
\label{ss:1.5D_line_formation_discussion}

Following \citet{Carlsson:1997}, we decomposed the contribution function $C_\mathrm{I}$ into three components in Eq.~\ref{eq:contribution_function}. 
This formalism quantifies the interplay of velocity gradients, the source function, and the \textit{remaining} optical thickness of the atmosphere on the emergent spectrum. 
For the transitions chosen in this study, we have only considered the properties of their line cores rather than those of the full emergent spectra. 
Since the first term, $\chi_\nu/\tau_\nu$, characterises the influence of velocity gradients we have not analysed this quantity in detail. 
Nevertheless, in the 1.5D \Lw{} approximation we find this quantity to peak for all line cores, and both projections, in the \ac{PCTR} closest to the observer. 
The second term of Eq.~\ref{eq:contribution_function}, $S_\nu$, has previously been introduced. 
Finally, the third component, $\tau_\nu e^{-\tau_\nu}$, peaks at $1/e$ where $\tau_\nu = 1$, a location overlaid on the $\tau_\nu$ column of Figure~\ref{fig:formation_properties_filament}. 

We find the source function for the filament projection to be highly structured in particular for those lines that are approximately optically thin \textit{i.e.}, \halpha{}\,--\,\Caii{~8542}, in comparison to their corresponding contribution function which appear rather broad along a given column. 
The inverse is then the case for the more optically-thick lines. 
Generally speaking, such relationships remain true for the prominence projection, but the thinner structures lead to several \ac{PCTR}s along the \ac{LOS} and hence also multiply-peaked contribution functions. 
Since the filament/prominence simulation used here is of relatively low resolution, the presence of individual threads are largely restricted to the prominence projection \citep[cf.][]{Xia:2016b, Jenkins:2022}.
The filament projection resembles instead a `monolithic' internal structure along the \ac{LOS}, cf. the $z$-projection in the right panel of Figure~\ref{fig:topology_example_filament_spectra}, that matches more closely the geometry of the earlier isothermal/isobaric/\ac{PCTR} models. 
We hence anticipate this difference in internal structuring, in terms of source and contribution functions (but also the primitive variables, cf. Figures~\ref{fig:average_formation_properties_filament}~\&~\ref{fig:average_formation_properties_prominence}), to be highly influenced by the coarsely resolved simulation used here. 
Since the \citetalias{Heinzel:2015} proxy synthesis of the higher resolution simulation of \citet{Jenkins:2022} yielded threaded appearances for the filament in addition to the prominence projection, we anticipate the equivalent \Lw{} analysis would be similarly highly structured \citep[][]{Heinzel:2006}.




The $C_\mathrm{I}$ associated measure of the \textit{average formation height} $z_\mathrm{a}$ quantity weighs the height so as to approximate the average height over which each of the constituent $C_\mathrm{I}$ components \textit{deposits} the majority of their information into the final emergent intensity. Alternatively, the Eddington-Barbier approximation assumes the height of formation to be where $\tau_\nu=1$ since photons emitted past this point (at higher $\tau_\nu$ $\equiv$ deeper depths) will likely be scattered/absorbed before reaching the observer \citep[][]{Vernazza:1981}. This approximation assumes there exists a location of $\tau_\nu=1$ along the \ac{LOS} \textit{i.e.}, the approximation is restricted to optically-thick atmospheres. In the filament projection case, it was found that $I_\nu \approx S_\nu(\tau_\nu=1) \approx S_\nu(\tau_\nu(z=z_\mathrm{a}))$, meaning there was a weak argument for favoring one approximation over the other. For the prominence projection, although not explicitly shown here, many of the columns do not contain $\tau_\nu=1$ yet show a clear emission signature in Figure~\ref{fig:prominence_comparison} nonetheless; the identical situation was found for some of the columns in the filament synthesis for the \halpha{} and \Caii{~8542} lines where the spectral properties were dominated by the background chromosphere. Thus, considering both the filament and prominence projections in addition to the potentially-wide range of $\tau_\nu$ encountered within an observed filament/prominence, we tend to agree with \citet{Leenaarts:2012a} in their preference for the $S_\nu(\tau_\nu(z=z_\mathrm{a}))$ approximation over $S_\nu(\tau_\nu=1)$.

The comparison between the locations of average formation height $z_\mathrm{a}$ for each line core, and the thermodynamic quantities in Figure~\ref{fig:average_formation_properties_filament}~\&~\ref{fig:average_formation_properties_prominence} provide us with a first order estimate for how deep within a filament/prominence the line cores are formed. In terms of Figures~\ref{fig:filament_comparison}~\&~\ref{fig:prominence_comparison}, this enables an understanding of which structures within the filament/prominence we are actually seeing. The broadest, darkest filament appearance is for the \Mgii{~h\&k} synthesis, explained with $z_\mathrm{a}$ as the \ac{LOS} getting stuck in the outermost layers. Each of the \Caii{} line cores describe similar fine structuring, the lowest-forming being \Caii{~8542} with a very weak absorption signature donated from only the densest (column mass) locations, whereas the higher-forming H\&K lines are contributed to by the hotter, less dense regions in between. \halpha{} then has a relatively low-lying $z_\mathrm{a}$ formation height, but with a contribution both large in magnitude and broad in extent throughout the filament, yet additionally peaked in the upper chromosphere suggesting an absorption signature influenced by the majority of the atmosphere rather than any specific position. This gives rise to the appearance of a broad absorption signature containing similar fine structures as in the \Caii{} lines. The ordering of these formation heights within the filament are in accordance with those found by \citet{Bjorgen:2019} for a model of an active-region chromosphere. For our filament atmospheres, however, the formation heights of these lines span a range of 10~Mm. For the prominence projection, the range of $z_\mathrm{a}$ is generally narrower on account of the thinner structure, and similarly lower $\tau_\nu$ encountered by each \ac{LOS}. In those locations where the \ac{LOS} traverses multi-threaded conditions, on the otherhand, $z_\mathrm{a}$ varies once more throughout the column by $\approx$~10~Mm. Hence, the appearance of the prominence in the synthesis of Figure~\ref{fig:prominence_comparison} is relatively uniform except in the presence of multiple threads where one can instead identify features that are visible in one line core but not another. For example, the signature of the strong contribution function for \halpha{} at $z=0$, $x=-7$~Mm in Figure~\ref{fig:formation_properties_prominence}, and co-located $z_\mathrm{a}=-7$~Mm in Figure~\ref{fig:average_formation_properties_prominence}, then explains how the bright feature at $z=0$, $y=14$~Mm of Figure~\ref{fig:prominence_comparison} is visible only in the \halpha{} panel (cf. the equivalent for \Mgii{~h\&k}). We will discuss the implications of such a wide range of formation heights in the following section.

\subsection{Appearance of the filament and prominence: limitations and outlook} \label{ss:limitations_and_future}

\subsubsection{Boundary condition}

The radiation pumping detailed in Section~\ref{sss:15d_two-part_comparison} was found to be a consequence of the assumed 1.5D infinite horizontal extent within our \Lw{} model geometry that meant radiation emitted from the chromosphere was unable to `free-stream' out of the system \citep[][]{Paletou:1993}.
In contrast, filaments in observations are characterised by their finite width and comparably long extent. 
As such, one of the directions that the 1.5D geometry assumes to be invariant maintains this property, at least approximately, for a filament within the actual solar atmosphere.
Adopting instead a 2.5D geometry \citep[as in][]{Paletou:1993}, the finite width of the filament in one of the dimensions would permit the `free-streaming' of radiation and prevent the significant iterative pumping we have shown in Figure~\ref{fig:15d_vs_2part}.
From this it seems probable that any attempt at a 1.5D \ac{NLTE} synthesis of an atmospheric stratification that includes both a chromosphere and a filament, even self-consistently, will struggle to reproduce realistic spectral line properties \textit{e.g.}, intensity, as we found here \citep[see also, ][]{Xia:2016b, Zhao:2017, Zhao:2019, Diazbaso:2019}. 
This likely explains why some features within the aforementioned 1.5D chromospheric models yield significantly different line core intensities when compared against an equivalent 3D synthesis \citep[typically described as remedied by some spatial source function smoothing in 3D][]{Leenaarts:2012a}. Specifically, the common reference to a persistent granular pattern imprinted upon those fibrils synthesised in 1.5D may be a consequence of a similarly-enhanced source function at granulation heights far below the fibril as we have shown for the stacked-atmosphere filament in Figure~\ref{fig:15d_aniso_temperature}.

Moving the FAL-C component of the \Lw{} model into the boundary was necessary to yield comparably negative-contrast line cores for filament atmospheres, however this approach artificially prohibits \textit{all} response in the chromosphere to the overlying filament.
It is not immediately clear \textit{how much} influence a real-world filament has on its underlying atmosphere.
Since one of the directions under the 2.5D geometry assumption remains invariant, radiation trapping will still occur for a model that contains both a chromosphere and a filament. 
It is likely that this mechanism will then contribute similar, but presumably reduced, enhancements to the line core intensity in the vicinity of the filament \citep[in fact this is already strongly indicated by ][to be a contribution directly from the chromosphere below]{Paletou:1993}.
This may be very relevant to the ongoing discussion on the `bright-rim' filament/prominence phenomenon commonly remarked upon within observations \citep[see,][]{Heinzel:1995, Paletou:1997}.

\subsubsection{1.5D Geometry}

The 1.5D statistical equilibrium and radiative transfer calculations within \Lw{} are carried out separately on each of the 1D stratified columns extracted from the \MpiAmr{} filament/prominence simulation. 
Hence, the influence of adjacent columns, be them directly neighboring or comparably distant, on any single 1D stratification is entirely neglected. 
For the filament projection it is, therefore, not possible to consider how lateral chromospheric radiation, incident on the sides of the entire filament body, may reach into and alter the statistical equilibrium of a central column of plasma.
In fact, the slice through the filament projection shown in Figures~\ref{fig:formation_properties_filament}~\&~\ref{fig:average_formation_properties_filament} was specifically chosen to trace the filament spine as it is a feature approximately furthest from the vertical edges of the filament, in the horizontal direction (see the `Z projection' of Figure~\ref{fig:topology_example_filament_spectra}).
As such, for our setup this cut represents the location within the filament presumed to be the \textit{least} influenced by the neglected lateral radiation incident on the side of the filament - we previously described the columns within this cut as being most valid in terms of the 1.5D approximation. 
However, this is already shown to be a liberal assumption by the equivalent prominence synthesis in Figures~\ref{fig:formation_properties_prominence}~\&~\ref{fig:average_formation_properties_prominence}.
That is, with the exception of the very optically-thick \Mgii{~h\&k} lines, the contribution function describes a smooth increase throughout the centre of the prominence atmospheres.
Those contribution functions presented in Figure~\ref{fig:average_formation_properties_prominence} are, however, calculated for an observer at the edge of the domain. 
To ascertain how much influence there is of lateral radiation on the spine of the filament/prominence, the contribution quantity should be instead evaluated in both directions towards the spine position.
The same argumentation is valid for the question of escape probability for those photons present within these regions central to the filament/prominence. In general, however, photons with trajectories more oblique to the vertical within a 1.5D atmosphere can be artificially trapped here, potentially contributing an enhancement of the local source and contribution functions.
Although not explicitly quantified, these figures are sufficient to indicate that the lateral energy balance remains non-negligible to first order in the locations we have taken to trace the filament spine, and so it appears clear that moving forward a geometry of at least 2.5D is to some degree necessary \citep[][]{Heinzel:2005, Gunar:2007a}.
Similarly for the prominence projection, neither radiation directly incident on the underside of the prominence, nor the attenuated contribution of this radiation as it passes through those layers of the prominence in between, was considered. 
The influence this has on the emergent specific intensity is anticipated to be of equal importance to that indicated here for the filament projection.

Recent decades have seen numerous works demonstrate and quantify the influence of the 1.5 versus 3D approximations on a range of spectral lines formed within the solar chromosphere \citep[\textit{e.g.},][]{Uitenbroek:1989a,Leenaarts:2009a,Leenaarts:2012a, Leenaarts:2013a,Bjorgen:2018}. 
We may assume, as for the chromosphere, that switching to a 2.5D representation will not influence the optical thickness encountered along a given \ac{LOS} for a specific wavelength \citep[][]{Leenaarts:2010}. 
Hence, even in 3D, the measured specific intensity will remain approximately governed by the local properties of the outer-most, $z(\tau_\nu \approx 1)\approx z_\mathrm{a}$ layer of the filament \textit{i.e.}, the \ac{PCTR}, as we have already seen for the \Mgii{~h\&k} line cores \citep[a feature previously remarked upon within the 2.5D \ac{PCTR} modelling of][]{Labrosse:2016}. 
The energy considerations within the statistical equilibrium atmosphere \textit{will}, however, differ.
This is clearly of importance as the source function is largely set by the non-local radiation field that will, in 2.5 or 3D, have access to additional spatial pathways. 
In 3D, this has been shown by \citet{Leenaarts:2012a}, \citet{Leenaarts:2013a}, and \citet{Bjorgen:2018} to smooth the source function and, given the opacity response, smooth the appearance of structures in the synthesis - repeatedly so for the chromospheric fibril phenomenon. 
Indeed, this smoothing is expected for the filament projection of each of the lines we have shown here.

As already discussed, the notably dark appearance of the filament body according to the optically thick \Mgii{~h\&k}, and similarly the \Caii{~H\&K}, resonance lines is a consequence of the radiation incident directly from the solar disk being entirely masked.
We know the measured intensity within the filament body for \Mgii{~h\&k} is heavily dependent on the source function, which is in turn equal to the average radiation field $J$ since $\tau_\nu$ is so large (\ac{EB} approximation). \citet{Leenaarts:2013a} and \citet{Bjorgen:2018} demonstrated a secondary influence of the 3D radiation field to be a decrease in contrast, along with an increase in average brightness. Whereas the latter can be attributed to a combination of the aforementioned smoothing and additional angles of photon escape, the former is a clear demonstrator of lateral radiation enhancing the local source function along any given \ac{LOS}. We therefore anticipate reduced contrast within the core of our filament when synthesised in higher dimensions.

We have so far highlighted how a multi dimensional geometry is necessary as it will influence the line core line depth and in turn the overall contrast of the 2D images in Figure~\ref{fig:filament_comparison}~\&~\ref{fig:prominence_comparison}. Given the additional complexity of line formation within the wings of the studied spectral lines, we have currently avoided any attempts at drawing conclusions based on these portions of the profiles. It is for this reason that we have focused purely on the line core components of the \MpiAmr{} simulation synthesis in this manuscript.

\section{Summary and Conclusions}\label{s:summary_and_conclusions}

We have introduced here the possibility to synthesise accurate line core contrasts according to 1.5D \ac{NLTE} statistical equilibrium and radiative transfer calculations for solar prominence/filament atmospheres extracted from a self-consistent \MpiAmr{} \ac{MHD} simulation.
We have demonstrated how the base 1.5D approximation struggles to handle atmospheres including both a chromosphere and elevated filament, overcome here by including new and suitable boundary conditions into the modular \Lw{} framework.
Both the individual pixel-by-pixel spectra, and 2D line core contrast images describe filament/prominence projections consistent with observations; the \Lw{} framework approach is found to yield close agreement with the state-of-the-art \halpha{} approximate method of \citetalias{Heinzel:2015}. 
Decomposing the terms of the \ac{RTE} into the components of the contribution function, we show how the self-consistently generated fine structures of the simulated filament/prominence impart intricate distributions within the line core formation properties of \halpha{}, \Caii{~H,~K,~\&~8542}, and \Mgii{~h\&k}.
By quantifying the associated average formation height, we find estimations consistent to that of the \ac{EB} approximation for filament atmospheres, whilst additionally enabling such estimations for prominences.
Looking forward, it is clear that a higher dimensional synthesis approach is needed.
As a 2.5D approach will already remove the restriction of studying the formation properties of the line core alone, it represents the natural next step in the march to full 3D.

\begin{acknowledgements}
We are indebted to Petr Heinzel, Stanislav Gunar, Brigitte Schmieder, and Ivan Mili\'{c} for the many lengthy conversations. We also thank the referee for their constructive comments. We acknowledge the open source software that made possible the data visualisations presented within this work (\href{https://www.python.org}{Python};\href{https://yt-project.org}{yt-project};\href{https://matplotlib.org}{matplotlib}). RK and JMJ are supported by the ERC Advanced Grant PROMINENT and an FWO grant G0B4521N. This project has received funding from the European Research Council (ERC) under the European Union's Horizon 2020 research and innovation programme (grant agreement No. 833251 PROMINENT ERC-ADG 2018). This research is further supported by Internal funds KU Leuven, project C14/19/089 TRACESpace. The computational resources and services used in this work were provided by the VSC (Flemish Supercomputer Center), funded by the Research Foundation Flanders (FWO) and the Flemish Government - department EWI.
CMJO acknowledges support from the University of Glasgow's College of Science and Engineering.
\end{acknowledgements}


\bibliographystyle{aa} 
\bibliography{bibliography,CmoRefs} 

\begin{appendix}
\section{\Lw{} Geometry and Boundary Conditions for Filament/Prominence Atmospheres}
\label{a:boundaries}


\begin{figure*}
    \centering
    \resizebox{0.65\hsize}{!}{\includegraphics[trim = 50 170 20 150, clip=]{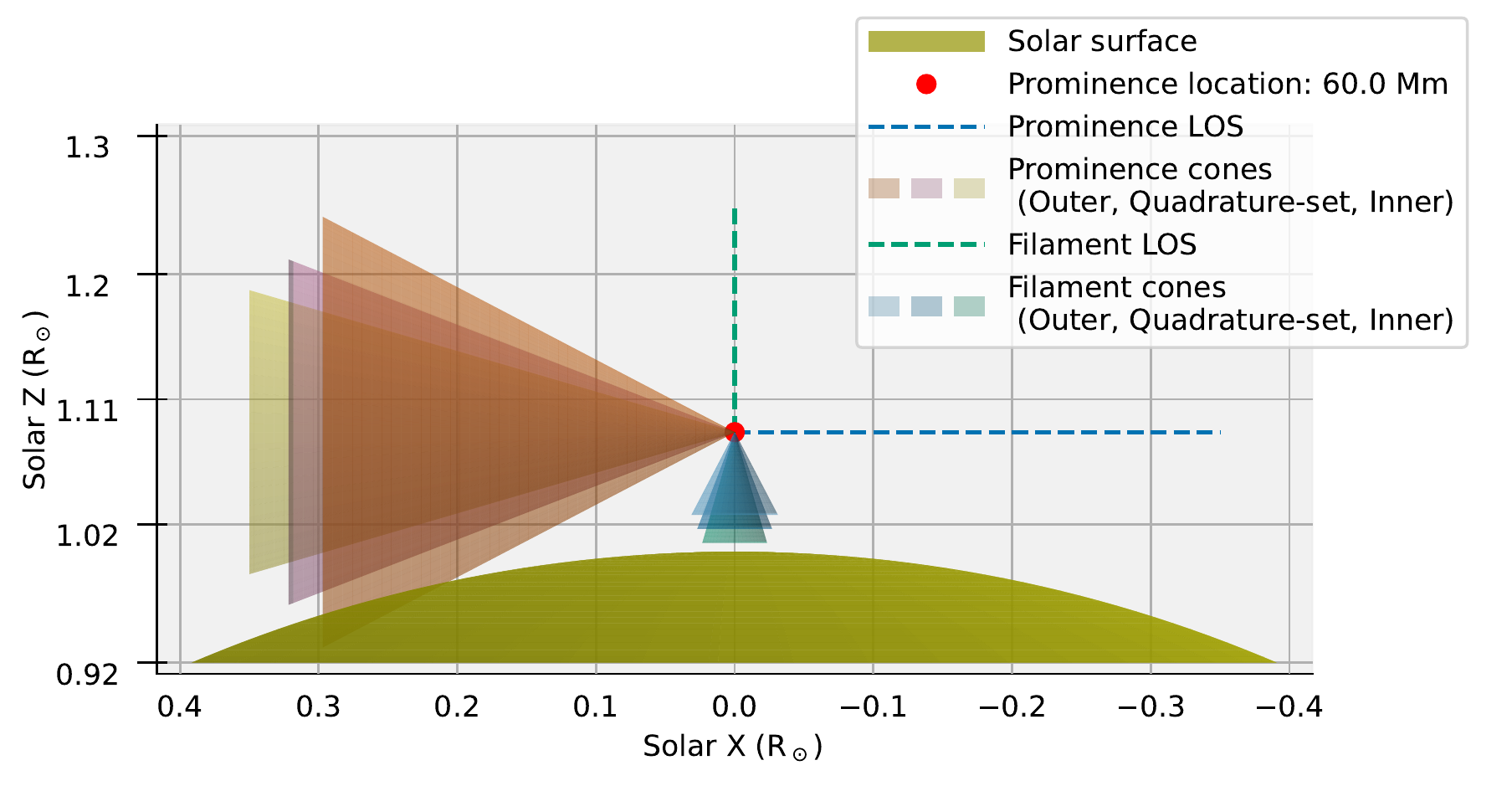}}
    \resizebox{0.72\hsize}{!}{ 
    \includegraphics[trim= 10 375 520.5 0, clip=]{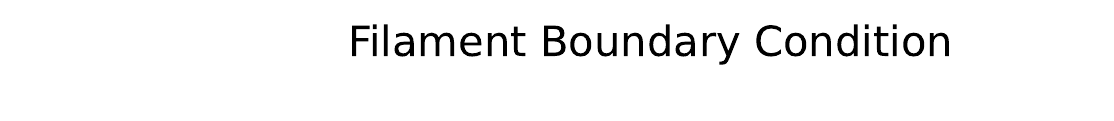}
    \hspace{-0.21cm}
    \includegraphics[trim= 520.2 375 20 0, clip=]{Figures/fil_prom_BCs_top.pdf}
    }
    \resizebox{0.72\hsize}{!}{ 
    \includegraphics[trim= 10 0 520.5 80, clip=]{Figures/fil_prom_BCs_top.pdf}
    \hspace{-0.21cm}
    \includegraphics[trim= 520.2 0 20 80, clip=]{Figures/fil_prom_BCs_top.pdf}
    }
    \caption{The approach for handling both filament and prominence boundary conditions within \Lw{}. \textit{Top panel}; the cone-averaging approach employed to ensure proper sampling of the limb darkening function with height. \textit{Bottom panels}; the tracing of the base quadrature across the solar surface for both the filament (left) and prominence (right) cases. These panels consider a view top-down onto the solar surface from \textit{e.g.}, Solar~Z = 1.3~$R_\odot$. This base quadrature sampling is augmented with the additional cones as shown in the top panel.}
    \label{fig:fil_prom_BCs}
\end{figure*}


The default geometry of a 1.5D \Lw{} \texttt{Atmosphere} python object orients the provided atmosphere, necessarily plane-parallel, along the $z$ direction and solves the energy transport for a quadrature of rays in both the $z$ and $x$ directions. Specifically, the quadrature is defined in the $z$~direction for $0<\mu_z<1$, where $\mu_z=\mathrm{cos}~\theta_z$ hence $0\,<\theta_z<\,\pi/2$. In a plane-parallel model, we assume the atmosphere only varies with $z$, so there's a cylindrical symmetry term that lets us drop any $y$ dependence (by convention, could be $x$ without loss of generality). Any form of symmetry breaking requires a more complex quadrature. The subsequent statistical equilibrium assumes that the angular transport of energy is 2D $\mu_z$\,--\,$\mu_x$ and well-resolved using a low fifth-order \ac{GL} quadrature. This yields 10 total rays when considering both the upward and downward propagation of energy, as shown in Figure~\ref{fig:15d_aniso_temperature}. This quadrature geometry is then consistent throughout both the atmospheric stratification and associated boundary conditions. A more detailed discussion of the internal machinery of \Lw{} is available in \cite{Osborne:2021}. Our treatment of the radiation within these boundary conditions then differs from the default ThermalisedRadiation + ZeroRadiation combination by considering the embedding of the empirical, chromospheric FAL-C atmosphere within the bottom boundary as discussed in Sections~\ref{sss:1.5D_geom}\,--\,\ref{sss:15d_two-part_comparison}. In the following, we will detail how we utilise this modified boundary condition in combination with the \MpiAmr{} atmosphere columns so as to yield the synthesised filament/prominence projections of Section~\ref{s:Results}.

\subsection{Filament Model Boundary Conditions}\label{a:filament_boundaries}

For a vertical model filament, the normal to the solar surface is parallel to the filament-bottom normal and hence a plane-parallel atmosphere extracted along the negative~$y$ axis of the \MpiAmr{} simulation cube. The $z$-axis then exactly aligns between the boundary conditions and filament atmosphere and the associated quadrature is shared identically, as is a default assumption for a \Lw{} atmosphere. As detailed in Section~\ref{sss:two-part_geom}, the two-part boundary condition contains a ThermalisedRadiation + FAL-C atmosphere that is synthesised \textit{a-priori} for 100 $\mu_z$-angles. The angle $\mu_z$ and wavelength $\lambda$ dependent radiation incident on the underside of the filament is obtained by equating the input radiation for a given quadrature angle $\mu_z$ according to,
\begin{equation}
    I_\mathrm{fil}^\mathrm{in}(\lambda,\mu_z) = I_\mathrm{chro}^\mathrm{out}\left(\lambda, 1.0 - (1.0-\mu_z^2) \frac{(R_\odot + h)^2}{R_\odot^2}\right), \label{eq:chromoI_from_muz}
\end{equation}
where $I_\mathrm{chro}$ is the specific intensity of the \textit{a-priori} ThermalisedRadiation + FAL-C atmosphere, and $h$ is the height of the filament above the solar surface at disk centre. Then, by assuming a spherically-symmetric solar surface and a fixed chromospheric illumination across this solar surface \textit{i.e.}, the FAL-C model, the incident radiation on the bottom of the filament is similarly spherically-symmetric. The bottom-left panel of Figure~\ref{fig:fil_prom_BCs} shows the quadrature sampling of the lower boundary condition, represented here as a \textit{solar} surface, according to the fixed quadrature for a filament at a height of 25~Mm - the rotational symmetry about the vertical $z$ axis is clear.

As the filament rises, the \textit{apparent} solar limb reaches farther away from the filament and the quadrature sampling adapts accordingly. In Section~\ref{sss:1.5D_geom}, it was discussed that as the height of the filament increases a fixed quadrature ray may no longer intersect the solar surface and be \textit{infinitely} limb-darkened. It was noted, however, in our early testing that the evolution in synthesised spectral intensity for an increasing height would not evolve in a smooth manner as anticipated. Although the chosen sampling of the incident radiation uses \ac{GL} quadrature and associated weights, this is due to the underlying assumption that a sufficient quadrature order represents a \textit{good sampling} of a function between two bounds. However, since we do not integrate the limb darkening function, once a ray no longer intersects the solar surface the incident radiation loses information about the limb-darkening profile close to the limb. This is overcome by instead averaging the radiation around a hollow cone of finite thickness, the centre of which is set by the \ac{GL} quadrature. Here, the opening angle and rotation are assumed to be well sampled by an additional \ac{GL} and trapezoidal quadrature, respectively. Hence, the cone-averaged ThermalisedRadiation + FAL-C specific intensity around each $\mu_z$ of the \ac{GL} quadrature is supplied to the filament and overcomes the aforementioned jumps in intensity for changes in height. An example quadrature for the filament projection is shown in the top panel of Figure~\ref{fig:fil_prom_BCs} in addition to the bounding cones employed for the angular average. For the case at hand we maintain the aforementioned spherical symmetry assumption as the chromospheric illumination is provided in our boundary conditions by the angular-dependent, but spatially constant, FAL-C model. Nevertheless, the current implementation generalises for the optional consideration of a spatially varying boundary condition.

\subsection{Prominence Model Boundary Conditions}\label{a:prominence_boundaries}

A \ac{LOS} intersecting an identically-vertical prominence model will be necessarily rotated by 90$^\degree$ compared to the filament model on account of the structure being located exactly at the solar limb from the perspective of any observer. Columns are thus extracted from the \MpiAmr{} simulation cube along the negative~$x$ axis and similarly set aligned with the $z$ coordinate within \Lw{}. This ensures that the quadrature in $\mu_z$ accurately transports the energy throughout the prominence atmosphere. The boundary conditions then handle the change in reference frame since the radiation from the solar surface is now incident on both the upper and lower ends of the prominence atmosphere, and along $\mu_x$ rather than $\mu_z$.

For a prominence projection, every horizontal row of extracted columns from the \MpiAmr{} simulation cube corresponds to a different height. This is in contrast to the filament case wherein a change in height is fixed for the entirety of a given synthesis. In this way, the influence of limb-darkening on the appearance of these 1.5D prominence atmospheres will be more pronounced. In addition, the cones of fixed $\mu_z$, for differing angles of rotation around the $z$~axis, no longer intersect the chromosphere at the same angle as was the case for the filament. This varying $\mu_x$ is then found simply using,
\begin{equation}
    \mu_x = \mathrm{cos}(\phi)\sqrt{1-\mu_z^2}, \label{eq:mux_from_muz}
\end{equation}
where $\phi$ measures the rotation around the $z$~axis. The bottom-right panel of Figure~\ref{fig:fil_prom_BCs} shows how a cone of fixed $\mu_z$ then traces across the solar surface, from which we can find the incident radiation intensity according to Eq.~\ref{eq:chromoI_from_muz} and swapping $\mu_z$ for $\mu_x$. Finally, we combine this modified boundary condition with the finite cone-averaging method described in Appendix~\ref{a:filament_boundaries} to ensure that the limb-darkening profile is properly sampled and provided to the prominence atmosphere.

\end{appendix}

\end{document}

%% file: acronyms.tex
\usepackage{acro}

\acsetup{
  single        = false,
  list/sort     = true,
  cite/group    = true,
  cite/group/cmd = \citealp,
  cite/group/pre = {; },
}

\DeclareAcronym{EVE}{
  short = EVE,
  long = \textit{Extreme Ultraviolet Variability Experiment},
  cite = {Woods:2012}}

\DeclareAcronym{AIA}{
  short = AIA,
  long = \textit{Atmospheric Imaging Assembly},
  cite = {Lemen:2012}}

\DeclareAcronym{HMI}{
  short = HMI,
  long = \textit{Helioseismic Magnetic Imager},
  cite = {Scherrer:2012}}
  
\DeclareAcronym{SDO}{
  short = SDO,
  long = \textit{Solar Dynamics Observatory},
  cite = {Pesnell:2012}}

\DeclareAcronym{STEREO}{
  short = STEREO,
  long = \textit{Solar Terrestrial Relations Observatory},
  cite = {Kaiser:2008}}

\DeclareAcronym{SECCHI}{
  short = SECCHI,
  long = \textit{Sun Earth Connection Coronal and Heliospheric Investigation},
  cite = {Howard:2008}}
  
\DeclareAcronym{SCIP}{
  short = SCIP,
  long = Sun Centered Imaging Package}
  
\DeclareAcronym{MGN}{
  short = MGN,
  long = multi-scale Gaussian normalisation,
  cite = {Morgan:2014}}

\DeclareAcronym{EUVI}{
  short = EUVI,
  long = EUVI}

\DeclareAcronym{CME}{
  short = CME,
  short-plural-form = CMEs,
  long = coronal mass ejection,
  long-plural-form = coronal mass ejections}
  
\DeclareAcronym{PIL}{
  short = PIL,
  short-plural-form = PILs,
  long = polarity inversion line,
  long-plural-form = polarity inversion lines}
  
\DeclareAcronym{LOS}{
  short = LOS,
  short-plural-form = LOS,
  long = line of sight,
  long-plural-form = lines of sight}
  
\DeclareAcronym{FOV}{
  short = FOV,
  short-plural-form = FOV,
  long = field of view,
  long-plural-form = fields of view}
  
\DeclareAcronym{EUV}{
  short = EUV,
  short-plural-form = EUV,
  long = extreme ultraviolet,
  long-plural-form = extreme ultraviolet}
  
\DeclareAcronym{GONG}{
  short = GONG,
  long = \textit{Global Oscillations Network Group}}
  
\DeclareAcronym{DST}{
  short = DST,
  long = \textit{Dunn Solar Telescope}}

\DeclareAcronym{IBIS}{
  short = IBIS,
  long = \textit{Interferometric Bidimensional Spectropolarimeter},
  cite = {Cavallini:2006,Reardon:2008}}

\DeclareAcronym{FIRS}{
  short = FIRS,
  long = \textit{Facility Infrared Spectropolarimeter},
  cite = {Jaeggli:2011}}
  
\DeclareAcronym{EIT}{
  short = EIT,
  long = \textit{Extreme Ultraviolet Imaging Telescope},
  cite = {Delaboudiniere:1995}}
 
\DeclareAcronym{ROSA}{
  short = ROSA,
  long = \textit{Rapid Oscillations in the Solar Atmosphere},
  cite = {Jess:2010}}
  
\DeclareAcronym{SPINOR}{
  short = SPINOR,
  long = \textit{Spectro-Polarimeter for Infrared and Optical Regions},
  cite = {Socas:2006}}  
 
\DeclareAcronym{SOHO}{
  short = SOHO,
  long = \textit{Solar and Heliospheric Observatory},
  cite = {Domingo:1995}}
 
\DeclareAcronym{MDI}{
  short = MDI,
  long = \textit{Michelson Doppler Imager},
  cite = {Scherrer:1995}}

\DeclareAcronym{NMSU}{
  short = NMSU,
  long = New Mexico State University}

\DeclareAcronym{NSO}{
  short = NSO,
  long = \textit{National Solar Observatory}}
  
\DeclareAcronym{TAC}{
  short = TAC,
  long = time allocation committee}
  
\DeclareAcronym{DEM}{
  short = DEM,
  short-plural-form = DEMs,
  long = Differential Emission Measure,
  long-plural-form = Differential Emission Measures}
  
\DeclareAcronym{PI}{
  short = PI,
  short-plural-form = PI,
  long = Principle Investigator,
  long-plural-form = Principle Investigators}  
  
\DeclareAcronym{IDL}{
  short = IDL,
  long = Interactive Data Language}
  
\DeclareAcronym{bb}{
  short = bb,
  long = broadband}  
  
\DeclareAcronym{nb}{
  short = nb,
  long = narrowband}    
  
\DeclareAcronym{IPM}{
  short = IPM,
  long = interplanetary medium}  
  
\DeclareAcronym{AO}{
  short = AO,
  long = adaptive optics,
  cite = {Rimmele:2004}} 
  
\DeclareAcronym{IR}{
  short = IR,
  long = infrared}  

\DeclareAcronym{halpha}{
  short = H-$\alpha$,
  long = Hydrogen-$\alpha$}

\DeclareAcronym{FPI}{
  short = FPI,
  long = Fabry-P\'erot}
  
\DeclareAcronym{DWDM}{
  short = DWDM,
  long = dense wavelength division multiplexing}  

\DeclareAcronym{CCD}{
  short = CCD,
  long = charge-coupled device}   
  
\DeclareAcronym{HI}{
  short = HI,
  long = Heliospheric Investigation}   
  
\DeclareAcronym{ROI}{
  short = ROI,
  long = region-of-interest}    
  
\DeclareAcronym{POV}{
  short = POV,
  long = point-of-view}  
  
\DeclareAcronym{MHS}{
  short = MHS,
  long = magnetohydrostatic}
  
\DeclareAcronym{MHD}{
  short = MHD,
  long = magnetohydrodynamic}  
  
\DeclareAcronym{RMHD}{
  short = RMHD,
  long = radiative magnetohydrodynamic}  
  
\DeclareAcronym{DOT}{
  short = DOT,
  long = \textit{Dutch Open Telescope},
  cite = {Rutten:1997}}  
  
\DeclareAcronym{BBSO}{
  short = BBSO,
  long = \textit{Big Bear Solar Observatory}}
  
\DeclareAcronym{NST}{
  short = NST,
  long = \textit{New Solar Telescope},
  cite = {Goode:2012}}

\DeclareAcronym{SOT}{
  short = SOT,
  long = \textit{Solar Optical Telescope},
  cite = {Tsuneta:2008}}
  
\DeclareAcronym{hinode}{
  short = Hinode,
  long = Hinode,
  cite = {Kosugi:2007}}
  
\DeclareAcronym{GST}{
  short = GST,
  long = \textit{Goode Solar Telescope}}  
  
\DeclareAcronym{SST}{
  short = SST,
  long = \textit{Swedish 1-m Solar Telescope},
  cite = {Scharmer:2002}}  
  
\DeclareAcronym{TRACE}{
  short = TRACE,
  long = \textit{Transition Region and Coronal Explorer},
  cite = {Handy:1999}}  
  
\DeclareAcronym{PCTR}{
  short = PCTR,
  long = \textit{prominence-corona-transition-region}}
  
\DeclareAcronym{DKIST}{
  short = DKIST,
  long = Daniel K. Inouye Solar Telescope}
  
\DeclareAcronym{RTI}{
  short = RTI,
  long = Rayleigh-Taylor instability}
  
\DeclareAcronym{SSW}{
  short = SSW,
  long = SolarSoftWare,
  cite = {Freeland:1998}}    
  
\DeclareAcronym{LTE}{
  short = LTE,
  long = local thermodynamic equilibrium}
  
\DeclareAcronym{NLTE}{
  short = NLTE,
  long = non local thermodynamic equilibrium}
  
\DeclareAcronym{FTS}{
  short = FTS,
  long = Fourier Transform Spectrometer,
  cite = {Kurucz:1984}}
  
\DeclareAcronym{RTE}{
  short = RTE,
  long = radiative transfer equation}
  
\DeclareAcronym{CRD}{
  short = CRD,
  long = complete frequency redistribution}
  
\DeclareAcronym{PRD}{
  short = PRD,
  long = partial frequency redistribution}
  
\DeclareAcronym{BCM}{
	short = BCM,
	long = Beckers' cloud model,
	cite={Beckers:1964}}
	
\DeclareAcronym{HSRA}{
	short = HSRA,
	long = Harvard Smithsonian Reference Atmosphere,
	cite={Gingerich:1971}}
	
\DeclareAcronym{NICOLE}{
	short = NICOLE,
	long = Non-LTE Inversion COde using the Lorien Engine,
	cite={Socasnavarro:2015}}
	
\DeclareAcronym{SIR}{
	short = SIR,
	long = Stokes Inversion based on Response functions,
	cite={Ruizcobo:1992}}
	
\DeclareAcronym{HAZEL}{
	short = HAZEL,
	long = Hanle and Zeeman Light,
	cite={AsensioRamos:2008}}
	
\DeclareAcronym{BPSS}{
	short = BPSS,
	long = bald patch separatrix surface,
	cite={Bungey:1996}}
	
\DeclareAcronym{EIS}{
	short = EIS,
	long = \textit{EUV Imaging Spectrometer},
	cite={Culhane:2007}}

\DeclareAcronym{FWHM}{
  short = FWHM,
  long = full width at half maximum}

\DeclareAcronym{AMRVAC}{
	short = MPI-AMRVAC,
	long = \textit{Adaptive Mesh Refinement Versatile Advection Code},
	cite = {Keppens:2012,Porth:2014,Xia:2018,Keppens:2020}}

\DeclareAcronym{AMR}{
	short = AMR,
	long = adaptive mesh refinement}

\DeclareAcronym{CCI}{
	short = CCI,
	long = Convective Continuum Instability}

\DeclareAcronym{BV}{
	short = BV,
	long = Brunt-V\"ais\"al\"a}

\DeclareAcronym{TVDLF}{
	short = TVDLF,
	long = Total Variation Diminishing Lax-Friedrich}

\DeclareAcronym{TI}{
	short = TI,
	long = Thermal Instability}

\DeclareAcronym{TNE}{
	short = TNE,
	long = thermal non-equilibrium}
	
\DeclareAcronym{MALI}{
    short = MALI,
    long = Multilevel Accelerated Lambda Iteration}
    
\DeclareAcronym{yt}{
    short = yt,
    long = yt-project,
    cite = {Turk:2011}}
    
\DeclareAcronym{GL}{
    short = GL,
    long = Gauss-Legendre}
    
\DeclareAcronym{CM}{
    short = CM,
    long = classically mounted}
    
\DeclareAcronym{MULTI3D}{
    short = MULTI3D,
    long = \textit{multi-level non-LTE 3D},
    cite = {Leenaarts:2009}}

\DeclareAcronym{EB}{
    short = EB,
    long = \textit{Eddington-Barbier}}
    
\DeclareAcronym{KDE}{
    short = KDE,
    long = kernel density estimate}
    
\DeclareAcronym{NLFFF}{
    short = NLFFF,
    long = nonlinear force-free field}